\documentclass{article}

\usepackage{arxiv}
\usepackage{mathrsfs}
\usepackage{cmap}

\usepackage[utf8]{inputenc} 
\usepackage[T1]{fontenc}    
\usepackage{hyperref}       
\usepackage{url}            
\usepackage{booktabs}       
\usepackage{amsfonts}       
\usepackage{nicefrac}       
\usepackage{microtype}      
\usepackage{graphicx}
\usepackage{doi}

\usepackage{amsthm}
\usepackage{stmaryrd}
\usepackage{cite}

\usepackage{times}
\usepackage{mathrsfs}
\usepackage{physics}
\usepackage[english]{babel}

\usepackage{mathtools}
\usepackage{indentfirst}
\usepackage{misccorr}
\usepackage{amssymb}
\usepackage[usenames]{color}
\usepackage{authblk}

\usepackage{graphicx}
\graphicspath{ {images/} }
\usepackage{amsmath}
\usepackage{float}
\usepackage{wrapfig}
\usepackage{comment}
\usepackage[export]{adjustbox}

\usepackage{geometry}
\title{Calculation of Some Integrals over Gaussian Measure with Nuclear Covariance Operator in Separable Hilbert Space}

\renewcommand{\shorttitle}{Calculation of Some Integrals over Gaussian Measure in Separable Hilbert Space}

\author[ ]{Nikita A. Ignatyuk$^{1,2,}$ \hspace{-0.5ex}\thanks{\href{mailto:Ignatyuk.na@phystech.edu}{Ignatyuk.na@phystech.edu}}\ }
\author[ ]{Anna A. Ogarkova$^{1,}$ \hspace{-0.5ex}\thanks{\href{mailto:Anna.a.ogarkova@mail.ru}{Anna.a.ogarkova@mail.ru}}\ }
\author[ ]{Stanislav L. Ogarkov$^{1,}$ \hspace{-0.5ex}\thanks{\href{mailto:Ogarkov.sl@phystech.edu}{Ogarkov.sl@phystech.edu}}}
\affil[1]{Moscow Institute of Physics and Technology (MIPT), Institutskiy Pereulok 9, 141701 Dolgoprudny, Russia}
\affil[2]{Skolkovo Institute of Science and Technology, Bolshoi Boulevard 30, bld. 1, 121205 Moscow, Russia}

\date{\vspace{-8ex}}

\begin{document}
\maketitle

\setcounter{page}{1}

\begin{abstract}

The main purpose of this paper is to construct convergent series for the approximate calculation of certain integrals over the Gaussian measure with a nuclear covariance operator, nonlocal propagator, in separable Hilbert space. Such series arise, for example, in the model with the interaction Lagrangian $\sinh^{2(p+1)}\varphi$, where $p \in \mathbb{N}$ and $\varphi$ is the scalar field, although the problem can be solved in general form for a fairly wide class of Lagrangians: an even, strictly convex, continuous, non-negative function with a single zero value for $\varphi=0$ and for $|\varphi|\rightarrow +\infty$ growing faster than $\varphi^{2}$. We strictly define the scattering matrix, $\mathcal{S}$-matrix, at the zero value of the classical field, argument of the $\mathcal{S}$-matrix, of such a theory in terms of the corresponding integral, find the iterated expansion for the integrand (the Gaussian measure doesn't expand) over two orthonormal bases of functions, prove the validity of the permutation of summation and integration and thus find the expansion of the $\mathcal{S}$-matrix at the zero value of the classical field into the iterated series in powers of the interaction action. In the process of deriving the interaction constant $g$ through an additional transformation can be transferred to the denominator. The individual terms of the resulting series have the form of a canonical partition function (CPF), and the methods of statistical physics are applicable to them. In particular, we express them in terms of Bell polynomials. It is important to note that such iterated series cannot be reduced to the perturbation theory (PT) series, since in the proposed model the latter diverges as $e^{n^{2}}$, where $n \in \mathbb {N}$ is the PT order. To study such series, one can use reasonable approximations for the arguments of Bell polynomials, for example, hard-sphere gas and stochastic clusters approximations. Along the way, we provide detailed mathematical background, including Beppo Levi's monotone convergence theorem (MCT) and Henri Lebesgue's dominated convergence theorem (DCT), without which the presented calculation would be significantly more complex.


\end{abstract}
\newpage
\tableofcontents
\section{Introduction}
\label{sect:introduction}

Nonlocal quantum field theory (NQFT) is a generalization of local quantum field theory (LQFT), based on the assumption of nonpoint interaction. This assumption is made in order to exclude ultraviolet divergences from the theory. The NQFT introduces a new parameter $l$ of the length dimension, which determines the size of the spacetime region where interactions occur. In the limit $l\rightarrow 0$ we return to the LQFT. In the case of one scalar field $\varphi$, and it is the case that we will consider in our paper, the typical NQFT Lagrangian has the form $\mathcal{L}(\varphi(x))=f(K(l\partial)\varphi(x))$, where the function $f$ depends on the result of the operator-valued function (form factor) $K(l\partial)$ action on the scalar field $\varphi$, that is, on the new scalar field $K(l\partial)\varphi(x)$. From the formal integration over all field configurations point of view, such a theory is equivalent to the theory with local Lagrangian $f(\varphi(x))$, but with modified by the form factor $K(l\partial)$ propagator, which can be proven by a formal linear change of fields $K(l\partial)\varphi(x)\rightarrow\varphi(x)$. Thus, when we talk about nonlocal quantum theory of a single scalar field, we are talking about a theory with nonlocal propagator or nuclear covariance operator.

Let us note that from the renormalization theory point of view, NQFT is a perturbatively, within the framework of perturbation theory (PT) with respect to the small coupling constant $g$, a nonrenormalizable theory. In the second half of the 20th century, there was an opinion that such theories should not exist at all. Fortunately, this opinion turned out to be wrong. To understand this, it is enough to recall, for example, the papers \cite{efimov1970nonlocal,efimov1977nonlocal,efimov1985problems,efimov1977cmp,efimov1979cmp}. An important advantage of such theories is that they are well defined in spacetime of any dimension $D$~\cite{efimov1970nonlocal,efimov1977nonlocal,efimov1985problems,efimov1977cmp,efimov1979cmp,basuev1973conv,basuev1975convYuk,ivanov1989confinement,ivanov1993quark,efimov2004blokhintsev,moffat1989,moffat1990,ekmw1991,moffat1991,moffat1994,moffat2011gravity,moffat2011higgs,moffat2016,alebastrov1973proof,alebastrov1974proof,Ogarkov2019I,Ogarkov2019II,Ogarkov2023,Modesto2014,Modesto2018,Modesto2015}.

With a nonlocal propagator, we can take the next step towards defining formal integration over all field configurations. Heuristically, the symbol $\int_{\varPhi}\mathcal{D}\left[\varphi\right]\,F\left[\varphi\right]$ is used to denote the integral of the functional $F\left[\varphi\right]$ with respect to ``measure'' $ \mathcal{D}\left[\varphi\right]$ in a space of all fields $\varPhi$. At the same time, the most important result of measure theory is that already in an infinite-dimensional Banach (complete normed) space (BS) non-trivial (non-zero) Lebesgue measure with all its properties doesn't exist. Therefore the measure $\mathcal{D}\left[\varphi\right]$ is not defined. Fortunately, both LQFT and NQFT always contain a free (Gaussian) theory (a theory without interaction). Therefore, having in our arsenal the nonlocal propagator, we can talk about the integral over the Gaussian measure specified by the free theory.

It remains to determine the space of all fields $\varPhi$, on whose Borel subsets the Gaussian measure will be specified. In this paper, we consider $\varPhi$ to be a Hilbert (complete with inner product) space (HS). This assumption is quite strong. So, for example, if we consider continuous functions as field configurations, $\varPhi$ will no longer be Hilbert. Moreover, depending on which set is the domain of fields, $\varPhi$ may not even be Banach. However, the main purpose of this paper is to develop methods for expanding integrals into series in powers of interaction action. Therefore, as a first step, the assumption that $\varPhi$ is HS seems natural. Selection of the appropriate space of field configurations and generalization of the results is the next independent problem. However, at the moment we can say that the framework of the BS turns out to be narrow. Apparently, a more suitable candidate for the space $\varPhi$ is a generic locally convex topological vector space (LCTVS)~\cite{Smolyanov}. In the general case, such a space is defined by a family of seminorms (there is no axiom of non-degeneracy) of arbitrary cardinality (not even necessarily countable), therefore, it is not even metrizable. However, the seminorms themselves are specified by the Minkowski functionals (distances to convex neighborhoods) of the LCTVS, that is, they are natural geometric constructions.

It is interesting to develop this idea in such a way that new seminorms arise with increasing energies at which quantum field processes are considered, that is the process of ``reverse compactification'' takes place. Such a process could lead to the ultraviolet scale $l$ change up to the observed, which would solve another fundamental problem of NQFT: the construction of a fundamental NQFT. As is known, NQFT is often used as an effective semiphenomenological theory. So, for example, the scattering matrix or $\mathcal{S}$-matrix of the Euclidean (the metric signature
is all pluses) NQFT with Lagrangians of a special form~\cite{efimov1970nonlocal,efimov1977nonlocal,efimov1985problems,efimov1977cmp,efimov1979cmp,basuev1973conv,basuev1975convYuk,Ogarkov2019I,Ogarkov2019II,Ogarkov2023,brydges1999review,rebenko1988review,brydges1980cmp,polyakov1987gauge,polyakov1977quark,samuel1978grand,o1999duality} is the grand canonical partition function (GCPF) of some gas with the interaction expressed by nonlocal propagator. This is how the duality between NQFT and statistical physics arises. Another application of NQFT is the method of functional (nonperturbative, exact) renormalization group (FRG)~\cite{vasil2004field,zinn1989field,kopbarsch,wipf2012statistical,rosten2012fundamentals,igarashi2009realization,Ogarkov2020III,Ogarkov2021IV}. The FRG method describes the flow of the NQFT family when the nonlocality scale $l$ changes to some LQFT, that is, for a fixed value of $l$ we have a nonlocal quantum theory. However, as far as we know, no fundamental nonlocal quantum theories have been found to date.

There is an extensive mathematical and physical literature devoted to Gaussian measures and functional integration~\cite{Mazzucchi2008,Mazzucchi2009,DeWitt-Morette,Kleinert,Johnson,Steiner,Simon,Shavgulidze2015,Bogachev,GiuseppePrato}. Many theorems on the localization of measures in LCTVS, as well as existence and uniqueness theorems for integrals over such measures have been proven. At the same time, the explicit calculation of integrals with respect to Gaussian measures is given for HS and less often BS, and the integrands are polynomials, linear and quadratic exponential functions, which corresponds to free NQFT. In this paper, our main purpose is to expand this ``list of integrals'' with more complex integrands. One may wonder whether such rigor within the framework of theoretical physics is necessary? The answer to this question is affirmative, primarily for the following reason. Within the framework of measure theory, many theorems that allow integration and calculation of the limit to be interchanged have been proven, including Beppo Levi's monotone convergence theorem (MCT) and Henri Lebesgue's dominated convergence theorem (DCT). Using these theorems, integrals over HS and BS can be represented in terms of the limit of finite multiplicity integrals. In general, such a representation doesn't always exist. Therefore, the theorems developed in measure theory make it possible not only to prove the nontriviality of a given integral in HS and BS, but also offers recipes for its calculation. The experience gained is important for the further research of constructive representations of integrals in generic LCTVS.

How one can obtain a convergent iterated series in theories with Lagrangians growing faster than the quadratic one? To explain the idea, let us consider the usual integral of the exponential function $e^{-m\varphi^{2}-g\varphi^{4}}$ (the parameters $m$ and $g$ are positive) with respect to one-dimensional Lebesgue measure $d\varphi$ on a real line $\mathbb{R}$, so $\varphi\in\mathbb{R}$. If we expand the exponential function $e^{-g\varphi^{4}}$ into a series and change summation and integration, then, as is easy to see, we arrive at an asymptotic series in powers of $g$. If we expand the exponential function $e^{-m\varphi^{2}}$ into a series and change summation and integration, we arrive at a convergent series in powers of $m$. The reason is that $e^{-g\varphi^{4}}$ is responsible for the convergence of the initial integral. In the case of the integral in HS, the situation is similar: the exponential function containing the interaction action is responsible for the convergence of the integral. Therefore, we need to find a way to ``hand over the responsibility for convergence'' to the Gaussian measure. To achieve this, it will be necessary to use methods of integrand expansion in a reasonable system of basis functions. Finding such a system is a fascinating mathematical problem, the solution of which requires knowledge of the orthogonal polynomials and special functions theory. If a suitable expansion is found, and it allows a permutation of summation and integration, we arrive at a certain series whose coefficients are integrals in HS, and in each integral the Gaussian measure is responsible for convergence. In this case, if we continue to select second expansion of integrand in each integral, now in terms of the interaction action, allowing for a rearrangement of summation and integration, we already arrive at an iterated series, the coefficients of which are the Gaussian averages of the interaction action powers, as in the case of PT. However, by construction, such an iterated series converges.

Why do we research the nonpolynomial interaction $\sinh^{2(p+1)}\varphi$, where $p\in\mathbb{N}$? As was mentioned above, the PT series in such a model will diverge faster than any factorial power, namely, as $e^{n^{2}}$, where $n\in\mathbb{N}$ is the PT order. For comparison: series for polynomial theories diverge as $e^{n\ln n}$, and such PT series can be summed using asymptotic methods, for example, Borel summation. Moreover, polynomial models can be obtained from the $\sinh^{2(p+1)}\varphi$ model in the small fields limit. For convenience, one can add an additional parameter to the hyperbolic sine argument. Finally, and this is also important, if it is possible to construct a series in terms of interaction action powers (perhaps with a modified propagator, but with a preserved functional dependence), different from the PT series, then the Gaussian averages of the interaction action powers are calculated explicitly in terms of Bell polynomials. This allows us to avoid the Isserlis--Wick combinatorics, combinatorics of Gaussian averages, which also makes the $\sinh^{2(p+1)}\varphi$ model very useful. The latter is a gift from statistical physics, and arises when calculating canonical partition function (CPF) of classical gas with interaction using the Mayer cluster expansion method.

In this paper we will focus on the calculation of $\mathcal{S}$-matrix. This is the central object of any quantum theory. And if it is possible to calculate it in convenient terms, then the theoretician’s problem is considered solved, and the rest is up to the experiment. And although the expression for the $\mathcal{S}$-matrix in terms of the limit of finite multiplicity integrals is mathematically the result, in QFT we want to obtain the result in terms of a series in (rational) powers of the coupling constant $g$. Let us also note that we are performing the calculation for a given Gaussian measure with nuclear covariance operator, so such a theory may depend on the ultraviolet scale. We don't consider any additional subtraction schemes. Our main purpose is to calculate the integral of the non-Gaussian exponential functional with respect to the Gaussian measure in separable HS in terms of a convergent series in powers of interaction action, and, for the convenience of numerical calculations, the coupling constant $g$ can be transferred to the denominator by some additional transformation in the process of constructing such a series. Thus, we propose an alternative to the asymptotic summation of PT series. A detailed calculation of each series term for various nonlocal propagators and subtraction schemes is beyond the scope of our paper.

The paper has the following structure. In the section~\ref{sect:gaussian-integrals-intro}, the general theory of Gaussian measures is considered, first in the finite-dimensional HS, and then, using the construction of a product measure, a generalization is constructed to the infinite-dimensional case, and definitions and theorems necessary for the calculation are given, the definition of $\mathcal{S}$-matrix at the zero value of the classical field (only this case is considered in the paper) is given and linear changes of variables are discussed. The section~\ref{sect:pert-theory} is devoted to the calculation of the $\mathcal{S}$-matrix by the PT method, for the terms of which, using the Mayer cluster expansion, the result is obtained in terms of Bell polynomials. In the section~\ref{sect:scattering-matrix}, main section of the paper, we find the $\mathcal{S}$-matrix in terms of the iterated series, the terms of which are again expressed in terms of Bell polynomials. Also in this section, the majorant and minorant for the $\mathcal{S}$-matrix are found for large values of the coupling constant $g$, which complements the expansion into the iterated series. In the \hyperref[sect:conclusion]{Conclusion}, general discussion of all the results obtained is given.
\section{Integrals over Gaussian Measure: Theoretical Introduction}
\label{sect:gaussian-integrals-intro}

\subsection{Basic Definitions and Notation from Measure Theory}

In this section we consider the general theory of Gaussian measures and also introduce the objects that we need in our paper using the books~\cite{Bogachev,GiuseppePrato}. Let a complete metric space $(X,\rho)$ ($\rho$ is a metric) be given, which in what follows we briefly denote by $X$. A system of subsets $\mathscr{A} \subset 2^{X}$ is called a $\sigma$-algebra if the following conditions are satisfied (without the last condition we obtain the definition of algebra):
\begin{enumerate}
    \item $X \in \mathscr{A}$.
    \item $\forall A \in \mathscr{A}$ and $\forall B \in \mathscr{A}$ it is true that $A \triangle B \in \mathscr{A}$ and $A \cap B \in \mathscr{A}$.
    \item $\forall (A_{n})_{n=1}^{\infty} \subset \mathscr{A}$ it is true that $\bigcup\limits_{n=1}^{\infty} A_{n} \in \mathscr{A}$.
\end{enumerate}

By $\mathscr{B}(X)$ we denote a Borel $\sigma$-algebra, that is $\sigma$-algebra generated by all open (equivalently, closed) subsets of $X$.

If some $\sigma$-algebra $\mathscr{A}$ of subsets of $X$ is given, then the pair $(X,\mathscr{A})$ is called a measurable space. We consider the measurable space $(X,\mathscr{B}(X))$

Let $\sigma$-algebra $\mathscr{A}$ and a set function $\mu: \mathscr{A}\rightarrow [0,+\infty]$ not identically equal to $+\infty$ be given. We say that $\mu$ is a measure on $\mathscr{A}$ if $\forall A \in \mathscr{A}$ and $\forall (A_{n})_{n=1}^{ N} \subset \mathscr{A}$ such that $A=\bigsqcup\limits_{n=1}^{N}A_{n}$, the equality $\mu(A)=\sum\limits_{n =1}^{N}\mu(A_{n})$ holds.

If, in addition, $\forall A \in \mathscr{A}$ and $\forall (A_{n})_{n=1}^{\infty} \subset \mathscr{A}$ such that $A =\bigsqcup\limits_{n=1}^{\infty}A_{n}$, the equality $\mu(A)=\sum\limits_{n=1}^{\infty}\mu(A_{n})$ holds, then the measure $\mu$ is called $\sigma$-additive. In what follows we consider only $\sigma$-additive measures. The most important example of such a measure is the Lebesgue measure on the real line $\mathbb{R}$.

If $\mu(X)<+\infty$, then such a measure is called finite. If $\mu(X)=1$, then such a measure is called probabilistic. A measure defined on $\mathscr{B}(X)$ is called Borel.

Let $(X,\mathscr{A})$ be a measurable space and a measure $\mu$ on $\mathscr{A}$ be given. The triple $(X,\mathscr{A},\mu)$ is called a measure space. In particular, if the measure is probabilistic, then the space is called probabilistic.

For each $A\in\mathscr{A}$, we denote by $A^{c}$ the complement of $A$, that is set $X\setminus A$. We use $\textbf{1}_A$ to denote a set indicator:
\begin{equation}
    \textbf{1}_A(x)=
    \begin{cases}
        1, & \mbox{if } x\in A, \\
        0, & \mbox{if } x\in A^c.
    \end{cases}    
\end{equation}

Let $\mathcal{H}$ be a real separable Hilbert space with scalar (inner) product $\langle\cdot|\cdot\rangle$ and norm $\Vert\cdot\Vert_{\mathcal{H}}^{}$. In this case, a measurable space $(\mathcal{H},\mathscr{B}(\mathcal{H}))$ and a space with Borel measure $(\mathcal{H},\mathscr{B}(\mathcal{H}),\mu)$ are defined.

We denote by $L(\mathcal{H})$ the linear space of all continuous linear operators from $\mathcal{H}$ to $\mathcal{H}$, by $L^{+}(\mathcal{H})$ the set of all $G \in L(\mathcal{H})$ symmetric, that is $\forall \varphi \in \mathcal{H}$ and $\forall \psi \in \mathcal{H}$ the equality $\langle G\varphi|\psi \rangle=\langle \varphi|G\psi\rangle$ holds, and positive ones, that is $\forall \varphi \in \mathcal{H}$ the inequality $\langle G\varphi|\varphi\rangle \geq 0$ holds.

Finally, we denote by $L_{1}^{+}(\mathcal{H})$ the set of all nuclear operators $G \in L^{+}(\mathcal{H})$, that is such operators for which the trace $\operatorname{Tr} G:=\sum\limits_{n=1}^{\infty}\left\langle G e_{n}|e_{n}\right\rangle < +\infty$ for one (and, therefore, for all) complete orthonormal system $\left(e_{n}\right)_{n=1}^{\infty} \subset \mathcal{H}$.

Let us note that if $G$ is a nuclear operator, then it is compact, that is maps any bounded subset in $\mathcal{H}$ to a relatively compact or precompact (closure is compact) subset in $\mathcal{H}$, and the trace of the operator $\operatorname{Tr} G$ is the sum of its eigenvalues repeated in according to their multiplicity (theorem from functional analysis).

\subsection{Gaussian Measures in Finite-Dimensional Hilbert Spaces}

As a next step, we consider the one-dimensional Hilbert space $\mathbb{R}$. In this case, for any pair $(a,\lambda)$ of real numbers $a \in \mathbb{R}$ and $\lambda \geq 0$ we define a probability measure $N_{a,\lambda}$ on $\mathscr {B}(\mathbb{R})$ as follows. If $\lambda=0$, we set $N_{a,0}=\delta_{a}$, where $\delta_{a}$ is the Dirac measure at $a$:
\begin{equation}
    \delta_a(A)=
    \left\{
        \begin{array}{ll}
        1, & \mbox{if } a \in A, \\
        0, & \mbox{if } a \notin A,
        \end{array} 
        \quad A \in \mathscr{B}(\mathbb{R}).
    \right.
\end{equation}

If $\lambda>0$, we give a definition in terms of the measure density
\begin{equation}
    N_{a,\lambda}(A)=\frac{1}{\sqrt{2\pi\lambda}}
    \int_{A} e^{-\frac{(x-a)^2}{2\lambda}} dx, 
    \quad A \in \mathscr{B}(\mathbb{R}).
\end{equation}

The measure $N_{a,\lambda}$ is a probability measure on $\mathscr{B}(\mathbb{R})$ due to the following equality, well known from analysis:
\begin{equation}
    N_{a,\lambda}(\mathbb{R})=
    \frac{1}{\sqrt{2\pi\lambda}} 
    \int_{\mathbb{R}} e^{-\frac{(x-a)^2}{2\lambda}} dx=1.    
\end{equation}

In the case of $\lambda>0$, the measure $N_{a,\lambda}$ is absolutely continuous with respect to the Lebesgue measure, that is for any null set (Lebesgue measurable set that has measure zero) the equality $N_{a,\lambda}(A)=0$ holds. In this case, we can formally set
\begin{equation}
N_{a,\lambda}(dx)=\frac{1}{\sqrt{2\pi\lambda}} 
e^{-\frac{(x-a)^2}{2\lambda}}dx.
\end{equation}

The quantities $a$ and $\lambda$ are called the mean and variance of the measure $N_{a,\lambda}$, respectively, as evidenced by the following easily verifiable equalities:
\begin{equation}
    \int_{\mathbb{R}} x N_{a, \lambda}(d x)=a, \quad
    \int_{\mathbb{R}}(x-a)^2 N_{a, \lambda}(dx)=\lambda.
\end{equation}

Finally, for the characteristic function $\widehat{N}_{a,\lambda}$ of the measure $N_{a,\lambda}$, that is for its Fourier transform, the following equality is true:
\begin{equation}
    \widehat{N}_{a,\lambda}(y):=\int_{\mathbb{R}} 
    e^{iyx}N_{a,\lambda}(dx)=e^{iay-\frac{1}{2}\lambda y^2}, 
    \quad y \in \mathbb{R}.  
\end{equation}

As a next step, we consider a Hilbert space of dimension $d \in \mathbb{N}$. Before defining Gaussian measures on $\mathscr{B}(\mathcal{H})$, let us recall some concepts about products of measures (product measures). We restrict ourselves to the case of finite measures, since Gaussian measures are finite (and even probabilistic).

Let for any $k=1,\ldots,d$ be given spaces with finite measures $\left(X_{k}, \mathscr{A}_{k}, \mu_{k}\right)$. Let us denote by $X$ the Cartesian product $X_1\times X_2\times\cdots\times X_d$. A subset $A \subset X$ is called a measurable rectangle if it has the form $A=A_1\times A_2\times\cdots\times A_d$, where $A_{k}\in\mathscr{A}_k$ for all $k =1,\ldots,d$.

The product of $\sigma$-algebras $\mathscr{A}_{k}$ over all $k=1,\ldots,d$ is the $\sigma$-algebra $\mathscr{A}$ generated by the set of all measurable rectangles.

For any measurable rectangle $A=A_1\times A_2\times\cdots\times A_d$ we define its measure $\mu(A)=\mu_{1}\left(A_1\right)\mu_{2}\left(A_2\right) \cdots\mu_{d}\left(A_d\right)$. Then the measure $\mu$ uniquely extends to a finite measure on $\mathscr{A}$, denoted by $\mu_{1}\times\mu_{2}\times\cdots\times \mu_{d}$. This statement has the most important theoretical significance: It guarantees not only the possibility, but also the correctness of constructing finite measures in large dimensions using finite measures in smaller dimensions.

Now let us return to the definition of Gaussian measures on $\mathscr{B}(\mathcal{H})$. This can be done for any vector $a \in \mathcal{H}$ and any operator $G \in L^{+}(\mathcal{H})$. Let $\left(e_1,\ldots,e_d\right)$ be an orthonormal basis in $\mathcal{H}$ of the operator $G$ eigenvectors, that is for any $k=1,\ldots,d$ the equality $G e_k=\lambda_k e_k, k= 1,\ldots, d$ holds for some $\lambda_k \geq 0$. For any $x \in \mathcal{H}$ and $k=1,\ldots,d$ we set by definition $x_k=\left\langle x|e_k\right\rangle$. This means we can identify $\mathcal{H}$ with $\mathbb{R}^d$ through the following isomorphism $\gamma$:
\begin{equation}
    \gamma: \mathcal{H}\rightarrow\mathbb{R}^d, 
    \quad x\rightarrow\gamma(x)=\left(x_1,\ldots,x_d\right).    
\end{equation}

Now let us define the probabilistic product measure $N_{a,G}$ on $\mathscr{B}\left(\mathbb{R}^d\right)$ as follows:
\begin{equation}
    N_{a,G}=N_{a_{1},\lambda_{1}}\times N_{a_{2},\lambda_{2}}
    \times\cdots\times N_{a_{d},\lambda_{d}}.    
\end{equation}

The measure defined in this way satisfies the following easily verifiable equalities, generalizing the one-dimensional case:
\begin{equation}
\begin{split}
    &\forall y \in \mathcal{H} \quad \int_{\mathcal{H}} 
    \langle y|x\rangle N_{a,G}(d x)=\langle a|y\rangle,\\
    &\forall y \in \mathcal{H} \quad \forall z \in \mathcal{H} \quad
    \int_{\mathcal{H}}\langle y|x-a\rangle
    \langle z|x-a\rangle N_{a,G}(d x)=\langle Gy|z\rangle.
\end{split}
\end{equation}

Thus, it is convenient (and even necessary for further generalization) to interpret the mean value (vector) $a$ and the covariance matrix (covariance operator or propagator) $G$ as linear and quadratic forms, respectively. In these terms, it is convenient to write the characteristic function $\widehat{N}_{a,G}$ of the measure $N_{a,G}$ in the following way:
\begin{equation}
    \widehat{N}_{a,G}(y):=
    \int_{\mathcal{H}}e^{i\langle y|x\rangle} N_{a,G}(d x)=
    e^{i\langle a|y\rangle -\frac{1}{2}\langle Gy|y\rangle}.
    \quad y \in \mathcal{H}.    
\end{equation}

As in the one-dimensional case, if the determinant of $G$ is positive, then the measure $N_{a,G}$ is absolutely continuous with respect to the Lebesgue measure in $\mathbb{R}^{d}$, which provides the following formal equality:
\begin{equation}
    N_{a,G}(dx)=\frac{1}{\sqrt{(2\pi)^{d}\operatorname{det}G}}\,
    e^{-\frac{1}{2}\left\langle G^{-1}(x-a)|x-a\right\rangle}dx.    
\end{equation}

The following statement is a consequence of the Fourier transform uniqueness. Let $\mathcal{H}$ be a finite-dimensional Hilbert space, vector $a \in \mathcal{H}$, operator $G\in L^{+}(\mathcal{H})$, measure $\mu$ -- a probability measure on $ \mathscr{B}(\mathcal{H})$ such that its characteristic function $\widehat{\mu}$ has the form::
\begin{equation}
    \widehat{\mu}(y)=
    \int_{\mathcal{H}}e^{i\langle y|x\rangle}\mu(dx)=
    e^{i\langle a|y\rangle -\frac{1}{2}\langle Gy|y\rangle}, 
    \quad y \in \mathcal{H}.    
\end{equation}

Then the measure $\mu=N_{a,G}$. This statement (the criterion that a measure is Gaussian) has a direct generalization in the infinite-dimensional case.

\subsection{Measures in Infinite-Dimensional Hilbert Spaces}

As a next step, we consider measures in the infinite-dimensional separable Hilbert space $\mathcal{H}$. Let the complete orthonormal system $\left(e_{n}\right)_{n=1}^{\infty} \subset \mathcal{H}$ be given. For any $n \in \mathbb{N}$ we define linear (continuous) projection operators or projectors $P_{n}: \mathcal{H} \rightarrow P_{n}(\mathcal{H})$ as follows:
\begin{equation}
    \forall \varphi \in \mathcal{H} \quad  
    P_{n} \varphi=\sum_{k=1}^{n}
    \left\langle \varphi|e_{k}\right\rangle e_{k}, 
    \quad \lim\limits_{n \rightarrow \infty} P_{n} \varphi=\varphi.
\label{Projection_operator}
\end{equation}

Convergence is determined by the norm $\Vert\cdot\Vert_{\mathcal{H}}^{}$. Next we define the space $M(\mathcal{H})$ of finite Borel measures on $\mathcal{H}$. The following statements are true, which we present without proof.

1) Let $\mu \in M(\mathcal{H})$ and $\nu \in M(\mathcal{H})$ such that the integrals are equal
\begin{equation}
    \int_{\mathcal{H}}f(\varphi)\mu(d\varphi)=
    \int_{\mathcal{H}}f(\varphi)\nu(d\varphi)
    \label{Equality_measures}
\end{equation}
for all continuous and bounded functions $f: \mathcal{H} \rightarrow \mathbb{R}$ on $\mathcal{H}$. Then the measures $\mu=\nu$.

2) Let $\mu \in M(\mathcal{H})$ and $\nu \in M(\mathcal{H})$ be probability measures. If $\forall \psi \in \mathcal{H}$ the equality of characteristic functions $\widehat{\mu}(\psi)=\widehat{\nu}(\psi)$ holds, then the measures $\mu=\nu$.

The characteristic function $\widehat{\mu}(\psi)$ is defined in a standard way:
\begin{equation}
    \widehat{\mu}(\psi):=
    \int_{\mathcal{H}} 
    e^{i\langle \varphi|\psi\rangle}\mu(d\varphi), 
    \quad \psi \in \mathcal{H}.    
\end{equation}

Let $\mu \in M(\mathcal{H})$ be a probability measure. Let us define the mean and covariance (do not confuse with variance) of $\mu$ using the experience of the finite-dimensional case. To determine the mean, we will assume that the integral
\begin{equation}
    \int_{\mathcal{H}}\Vert\varphi\Vert_{\mathcal{H}}^{}\,
    \mu(d\varphi)<+\infty.
\end{equation}

In this case $\forall\psi \in \mathcal{H}$ the linear functional $F: \mathcal{H} \rightarrow \mathbb{R}$, defined as
\begin{equation}
    F(\psi)=\int_\mathcal{H}\langle \varphi|\psi\rangle 
    \mu(d\varphi), 
\end{equation}
is continuous since the following majorizing estimate holds
\begin{equation}
    |F(\psi)| \leq \Vert\psi\Vert_{\mathcal{H}}^{}
    \int_\mathcal{H}\Vert\varphi\Vert_{\mathcal{H}}^{}\,
    \mu(d\varphi).    
\end{equation}

By the Riesz theorem on the representation of a linear continuous functional in $\mathcal{H}$ (Riesz representation theorem), there exists $m \in \mathcal{H}$ such that the following equality is true:
\begin{equation}
    \langle m|\psi\rangle=
    \int_\mathcal{H}\langle \varphi|\psi\rangle 
    \mu(d\varphi), \quad \psi \in \mathcal{H}.    
\end{equation}

The vector $m$ is called the mean of the measure $\mu$. The mean defines the (vector-valued) Pettis integral as follows:
\begin{equation}
    \int_\mathcal{H} \varphi \mu(d\varphi):=m.
\end{equation}

To determine the covariance of $\mu$ we will assume that the integral:
\begin{equation}
    \int_\mathcal{H}\Vert\varphi\Vert_{\mathcal{H}}^{2}\,
    \mu(d\varphi)<+\infty.  
\end{equation}

In this case $\forall\psi_{1} \in \mathcal{H}$ and $\forall\psi_{2} \in \mathcal{H}$ the bilinear form $Q: \mathcal{H} \times \mathcal {H} \rightarrow \mathbb{R}$, defined (using the vector $m$) as
\begin{equation}
    Q(\psi_{1},\psi_{2})=
    \int_\mathcal{H}\langle \psi_{1}|\varphi-m\rangle
    \langle \psi_{2}|\varphi-m\rangle \mu(d\varphi),   
\end{equation}
is continuous since the following majorizing estimate holds
\begin{equation}
    |Q(\psi_{1},\psi_{2})| \leq 
    \Vert\psi_{1}\Vert_{\mathcal{H}}^{}
    \Vert\psi_{2}\Vert_{\mathcal{H}}^{}
    \int_\mathcal{H}\Vert\varphi-m\Vert_{\mathcal{H}}^{2}\,
    \mu(d\varphi).    
\end{equation}

Therefore, again by Riesz representation theorem, there is a unique linear continuous operator $G \in L(\mathcal{H})$ such that $\forall\psi_{1} \in \mathcal{H}$ and $\forall \psi_{2} \in \mathcal{H}$ the following equality is true:
\begin{equation}
    \langle G\psi_{1}|\psi_{2}\rangle=
    \int_\mathcal{H}\langle \psi_{1}|\varphi-m\rangle
    \langle \psi_{2}|\varphi-m\rangle \mu(d\varphi).    
\end{equation}

The operator $G$ is called the covariance of $\mu$. It is also called covariance operator or propagator. The following statement is true, which we present without proof. Let $\mu \in M(\mathcal{H})$ be a probability measure with mean $m$ and covariance $G$. Then $G \in L_{1}^{+}(\mathcal{H})$, that is $G$ is a symmetric, positive and nuclear operator. Such propagators arise in NQFT.

\subsection{Gaussian Measures in Infinite-Dimensional Hilbert Spaces}

We are now ready to consider Gaussian measures in the infinite-dimensional separable Hilbert space $\mathcal{H}$. Let the vector $a \in \mathcal{H}$ and the operator and $G \in L_1^{+}(\mathcal{H})$ be given. Let us accept by definition that the Gaussian measure $\mu:=\mathcal{N}_{a,G}$ on $\mathscr{B}(\mathcal{H})$ is a measure $\mu$ having mean $a$, covariance operator $G$ and characteristic function $\widehat{\mathcal{N}}_{a,G}$ of the following form:
\begin{equation}
    \widehat{\mathcal{N}}_{a,G}(\psi)=
    \int_{\mathcal{H}} 
    e^{i\langle\varphi|\psi\rangle}
    \mathcal{N}_{a,G}(d\varphi)=
    \exp\left\{i\langle a|\psi\rangle-\frac{1}{2}\langle G\psi|\psi\rangle\right\}, \quad \psi \in \mathcal{H}.
\label{GM_Def}
\end{equation}

Gaussian measure $\mathcal{N}_{a,G}$ is called non-degenerate if the kernel $\operatorname{Ker}G$, that is the set of vectors $\{\varphi \in \mathcal{H}\,|\, G\varphi=0\}=\{0\}$. This definition is correct, that is for given $a \in \mathcal{H}$ and $G \in L_1^{+}(\mathcal{H})$ it defines a unique Gaussian measure $\mu=\mathcal{N}_{a,G}$ on $\mathscr{B}(\mathcal{H})$.

Let us note that since $G\in L_1^{+}(\mathcal{H})$, there are a complete orthonormal basis $\left(e_{n}\right)_{n=1}^{\infty} \subset \mathcal{H}$ of the operator $G$ eigenvectors and the sequence of non-negative numbers $\left(\lambda_{n}\right)_{n=1}^{\infty}$ (eigenvalues of the operator $G$) such that $\forall n \in \mathbb{N}$ the equality $G e_{n}=\lambda_{n} e_{n}$ holds.

Similar to the finite-dimensional case, for any $\varphi\in \mathcal{H}$ and $n \in \mathbb{N}$ we set $\varphi_{n}=\left\langle\varphi|e_{n}\right\rangle$. This gives rise to a natural isomorphism $\gamma$ between $\mathcal{H}$ and the Hilbert space $\ell^{2}$ of quadratically summable sequences $\left(\varphi_{n}\right)_{n=1}^{ \infty}$ of real numbers, that is sequences satisfying the following equality:
\begin{equation}
    \sum_{n=1}^{\infty}\left|\varphi_n\right|^{2}<+\infty.
\end{equation}

Natural isomorphism is defined as follows:
\begin{equation}
    \mathcal{H}\rightarrow \ell^{2},\quad 
    \varphi \rightarrow \gamma(\varphi)=
    \left(\varphi_n\right)_{n=1}^{\infty}.    
\end{equation}

By means of $\gamma$ we identify $\mathcal{H}$ and $\ell^{2}$ and consider the following measure in $\ell^{2}$:
\begin{equation}
    \mu:=\bigtimes\limits_{n=1}^{\infty} 
    N_{a_{n},\lambda_{n}}=N_{a_{1},\lambda_{1}}
    \times N_{a_{2},\lambda_{2}}
    \times\cdots
\end{equation}

At the moment this notation is formal, since we have yet to define such an infinite product of measures (infinite product measure), but for the sake of explanation of the final result it is useful to run past the relevant definitions and formulate the final result. At first glance, such a measure is defined in the space $\mathbb{R}^{\infty}:=\bigtimes\limits_{n=1}^{\infty}\mathbb{R}$ of all sequences of real numbers. However, due to the properties of the operator $G$ eigenvalues, this measure is actually concentrated in the space $\ell^{2}$, that is $\mu(\mathbb{R}^{\infty}\setminus\ell^{2})=0$. As it turns out, the measure $\mu$ is the Gaussian measure $\mathcal{N}_{a,G}$ (after identifying the spaces $\mathcal{H}$ and $\ell^{2}$). Let us proceed to defining the countable product of measures. We first discuss the case of arbitrary probability measures, and then consider the case of Gaussian measures in more detail.

\subsection{Countable Product of Measures}

Consider the linear space $\mathbb{R}^{\infty}$. Let us equip it with a metric $\rho$, thus turning it into a metric space. We define the metric $\rho$ as follows:
\begin{equation}
    \rho(\varphi,\psi)=\sum_{n=1}^{\infty}\frac{1}{2^{n}} 
    \frac{\max\left\{\left|\varphi_{{k}}-
    \psi_{k}\right||1\leq k \leq n\right\}}
    {1+\max\left\{\left|\varphi_{{k}}-
    \psi_{k}\right||1\leq k \leq n\right\}}, 
    \quad \varphi \in \mathbb{R}^{\infty},
    \quad \psi \in \mathbb{R}^{\infty}.
\end{equation}
The resulting metric space is complete. Moreover, the topology corresponding to the metric $\rho$ is exactly the Tychonoff topology (product topology).

Let $\left(\mu_{n}\right)_{n=1}^{\infty}$ be a sequence of probability measures on $\mathscr{B}(\mathbb{R})$. Let us define a countable product of measures $\mu:=\bigtimes\limits_{n=1}^{\infty}\mu_{n}$ on the family $\mathscr{C}(\mathbb{R}^{\infty})$ of all cylindrical subsets (cylinders) $C_{n,A} \subset \mathbb{R}^{\infty}$, where the parameters of the cylinder are $n \in \mathbb{N}$ and $A \in \mathscr{B }\left(\mathbb{R}^{n}\right)$, and the cylinder itself has the form:
\begin{equation}
    C_{n, A}=\left\{\varphi=\left(\varphi_k\right)_{k=1}^{\infty} 
    \in \mathbb{R}^{\infty}|\left(\varphi_1, \ldots, \varphi_n\right) 
    \in A\right\}. 
\end{equation}

For any $n \in \mathbb{N}$ and $k \in \mathbb{N}$ cylindrical sets satisfy the following equality (the index of $\mathbb{R}$ means the number of the factor in the Cartesian product, which is $\mathbb {R}$):
\begin{equation}
    C_{n,A}=C_{n+k,A\times\mathbb{R}_{n+1}\times\cdots\times\mathbb{R}_{n+k}}.
\end{equation}

The family $\mathscr{C}(\mathbb{R}^{\infty})$ of all cylindrical subsets of $\mathbb{R}^{\infty}$ is an algebra. Moreover, the $\sigma$-algebra generated by $\mathscr{C}(\mathbb{R}^{\infty})$ coincides with the Borel $\sigma$-algebra $\mathscr{B}\left(\mathbb{R}^{\infty}\right)$, since any ball (with respect to $\rho$ on $\mathbb{R}^{\infty}$) is a countable intersection of cylindrical sets.

Now we are ready to define the countable product of measures $\mu$. First we define $\mu$ on the algebra $\mathscr{C}(\mathbb{R}^{\infty})$. For any $C_{n,A} \in \mathscr{C}(\mathbb{R}^{\infty})$ we set
\begin{equation}
    \mu\left(C_{n,A}\right)=
    \left(\mu_{1}\times\mu_{2}\times\cdots\times \mu_{n}\right)(A).
\end{equation}

The measure $\mu$ defined in this way is a $\sigma$-additive measure on $\mathscr{C}(\mathbb{R}^{\infty})$ and uniquely extends to a $\sigma$-additive probability measure on $\mathscr{B}\left(\mathbb{R}^{\infty}\right)$. We present this statement without proof, although it is of utmost importance for the construction of probability measures in infinite-dimensional spaces.

\subsection{Countable Product of Gaussian Measures}

Now let us explain why the measure $\mu=\bigtimes\limits_{n=1}^{\infty} N_{a_{n},\lambda_{n}}$ is a Gaussian measure on $\ell^{2}$ with mean $a$ and covariance $G$. First of all, the set $\ell^{2} \in \mathscr{B}\left(\mathbb{R}^{\infty}\right)$, that is a Borel set. Let us define the Borel $\sigma$-algebra $\mathscr{B}\left(\ell^{2}\right)$ as the $\sigma$-algebra of sets of the form $\ell^{2} \cap A$, where $A \in \mathscr{B}\left(\mathbb{R}^{\infty}\right)$. It is clear that we can consider a restriction of the measure $\mu$ to $\mathscr{B}\left(\ell^{2}\right)$. And, since the equality $\mu(\mathbb{R}^{\infty}\setminus\ell^{2})=0$ is true, it is precisely this restriction that we will consider further.

Now let us formulate the main statement. There is a unique probability measure $\mu$ on $\mathscr{B}(\mathcal{H})$ with mean $a$, covariance $G$ and characteristic function $\widehat{\mu}$ of the following form::
\begin{equation}
    \widehat{\mu}(\psi)=
    \exp\left\{i\langle a|\psi\rangle-\frac{1}{2}\langle G\psi|\psi\rangle\right\}, \quad \psi \in \mathcal{H}.
\label{Theorem_1}
\end{equation}

We will denote such a measure by $\mathcal{N}_{a,G}$. If $a=0$, then instead of $\mathcal{N}_{0,G}$ we will write $\mathcal{N}_{G}$. To prove this, let us check that the restriction of the countable product of measures $\mu=\bigtimes\limits_{n=1}^{\infty} N_{a_{n},\lambda_{n}}$ to $\mathscr{B} \left(\ell^{2}\right)$ satisfies the required properties. In this case, the presence of natural isomorphism $\gamma$ will complete the proof.

In the proof, we will assume that the Gaussian measure is non-degenerate, that is $\operatorname{Ker}G=\{0\}$, and also (this is not a restriction) we order the eigenvalues of $G$ in a non-decreasing manner, that is $\lambda_1 \geq \lambda_2 \geq \cdots \geq \lambda_n \geq \cdots$.

First of all, the following equality is true:
\begin{equation}
    \int_{\ell^{2}}\Vert\varphi\Vert_{\ell^{2}}^{2}\,
    \mu(d\varphi)=\operatorname{Tr}G+
    \Vert a \Vert_{\ell^{2}}^{2}.
\end{equation}

Let $\left(P_{n}\right)_{n=1}^{\infty}$ be a sequence of projectors, and let $\psi \in \ell^{2}$. Since $|\langle \varphi|\psi\rangle| \leq \Vert \varphi \Vert_{\ell^{2}}\Vert \psi \Vert_{\ell^{2}}$ and $\int_{\ell^{2}}\Vert \varphi \Vert_{ \ell^{2}}\,\mu(d\varphi)$ is finite, then by Henri Lebesgue’s dominated convergence theorem (DCT) the following equality holds:
\begin{equation}
    \int_{\ell^{2}}\langle \varphi|\psi\rangle\,\mu(d\varphi)=
    \lim\limits_{n\rightarrow\infty} 
    \int_{\ell^{2}}\langle P_{n}\varphi|\psi\rangle\,\mu(d\varphi).
\end{equation}

Let us continue this equality with the following chain:
\begin{equation}
\begin{split}
    &\int_{\ell^{2}}\langle P_{n}\varphi|\psi\rangle\,\mu(d\varphi)=
    \sum\limits_{k=1}^{n}\int_{\ell^{2}}
    \varphi_{k}\psi_{k}\,\mu(d\varphi)=
    \sum\limits_{k=1}^{n}\psi_{k}\int_{\mathbb{R}}\varphi_{k}\, 
    N_{a_{k},\lambda_{k}}\left(d\varphi_{k}\right)=\\
    &\sum\limits_{k=1}^{n} \psi_{k} a_{k}=
    \left\langle P_{n}a|\psi\right\rangle\rightarrow
    \langle a|\psi\rangle=\int_{\ell^{2}}\langle \varphi|\psi\rangle\,\mu(d\varphi), \quad n \rightarrow \infty.
\end{split}
\end{equation}

Thus, the mean of $\mu$ is equal to $a$. Similarly, to determine the covariance of $\mu$, we fix $\psi_{1} \in \ell^{2}$ and $\psi_{2}\in \ell^{2}$ and write the following equality:
\begin{equation}
    \int_{\ell^{2}}\langle \varphi-a|\psi_{1}\rangle
    \langle \varphi-a|\psi_{2}\rangle\,\mu(d\varphi)=
    \lim\limits_{n\rightarrow\infty}
    \int_{\ell^{2}}\langle P_{n}(\varphi-a)|\psi_{1}\rangle
    \langle P_{n}(\varphi-a)|\psi_{2}\rangle\,\mu(d\varphi).
\end{equation}

Let us continue this equality with the following chain:
\begin{equation}
\begin{split}
    &\int_{\ell^{2}}\langle P_{n}(\varphi-a)|\psi_{1}\rangle
    \langle P_{n}(\varphi-a)|\psi_{2}\rangle\,\mu(d\varphi)=
    \sum\limits_{k=1}^{n}\int_{\ell^{2}}
    \left(\varphi_{k}-a_{k}\right)^{2}
    \psi_{1,k}\psi_{2,k}\,\mu(d\varphi)=\\
    &\sum\limits_{k=1}^{n}\psi_{1,k}\psi_{2,k}
    \int_{\mathbb{R}}\left(\varphi_{k}-a_{k}\right)^{2} 
    N_{a_{k},\lambda_{k}}\left(d\varphi_{k}\right)=
    \sum\limits_{k=1}^{n}\lambda_{k}\psi_{1,k}\psi_{2,k}=
    \left\langle P_{n}G\psi_{1}|\psi_{2}\right\rangle.
\end{split}
\end{equation}

So in the limit $n \rightarrow \infty$ we have:
\begin{equation}
    \left\langle P_{n}G\psi_{1}|\psi_{2}\right\rangle
    \rightarrow\left\langle G\psi_{1}|\psi_{2}\right\rangle=
    \int_{\ell^{2}}\langle \varphi-a|\psi_{1}\rangle
    \langle \varphi-a|\psi_{2}\rangle\,\mu(d\varphi)
\end{equation}

Thus, the covariance of $\mu$ is equal to $G$. Finally, for any $\psi \in \ell^{2}$ we have:
\begin{equation}
\begin{split}
    &\int_{\ell^{2}}e^{i\langle \varphi|\psi\rangle}\mu(d\varphi)=
    \lim\limits_{n\rightarrow\infty} 
    \int_{\ell^{2}}e^{i\langle P_{n}\varphi|\psi\rangle}\mu(d\varphi)
    =\lim\limits_{n\rightarrow\infty}
    \prod\limits_{k=1}^{n}
    \int_{\mathbb{R}}e^{i\varphi_{k}\psi_{k}}
    N_{a_{k},\lambda_{k}}\left(d\varphi_{k}\right)=\\
    &\lim\limits_{n\rightarrow\infty}
    \prod\limits_{k=1}^{n} 
    e^{i a_{k}\psi_{k}-\frac{1}{2}\lambda_{k}\psi_{k}^{2}}=
    \lim\limits_{n\rightarrow\infty} 
    e^{i\left\langle P_{n}a|\psi\right\rangle-
    \frac{1}{2}\left\langle P_{n}G\psi|\psi\right\rangle}=
    e^{i\left\langle a|\psi\right\rangle-
    \frac{1}{2}\left\langle G\psi|\psi\right\rangle}=
    \widehat{\mu}(\psi).
\end{split}
\end{equation}

Consequently, the characteristic function $\widehat{\mu}$ of the measure $\mu$ is determined by a Gaussian function. By identifying the spaces $\mathcal{H}$ and $\ell^{2}$, we obtain a unique probability measure $\mathcal{N}_{a,G}$ on $\mathscr{B}(\mathcal{H})$. Quod Erat Demonstrandum.

To complete the theoretical introduction on Gaussian measures in separable Hilbert spaces, let us calculate some integrals. The following parts of our paper are devoted to more complex integrals. To formulate the first example, we introduce the parameter $\varepsilon<\frac{1}{\lambda_1}$. For any such $\varepsilon$, the linear operator $1-\varepsilon\,G$ (by $1$ here we mean the identity operator) is invertible and the inverse operator $(1-\varepsilon\,G)^{-1}$ is bounded due to the following equality, valid for any $\varphi \in \ell^{2}$:
\begin{equation}
    (1-\varepsilon\,G)^{-1}\varphi=
    \sum\limits_{k=1}^{\infty}\frac{1}{1-\varepsilon\lambda_{k}}
    \left\langle \varphi|e_{k}\right\rangle e_{k}. 
\end{equation}

Therefore, in terms of an infinite product, we can define the determinant of the operator $(1-\varepsilon\,G)$ as follows:
\begin{equation}
    \operatorname{Det}(1-\varepsilon\,G):=
    \lim\limits_{n\rightarrow\infty}
    \prod\limits_{k=1}^{n}\left(1-\varepsilon\lambda_{k}\right):=
    \prod\limits_{k=1}^{\infty}\left(1-\varepsilon\lambda_{k}\right).    
\end{equation}

Since the operator $G$ is nuclear, the sum of its eigenvalues $\sum\limits_{k=1}^{\infty}\lambda_{k}<+\infty$, hence the infinite product, introduced above, is finite and positive.

Now let us formulate the example itself. Let $\varepsilon \in \mathbb{R}$. Then the equality holds (since the construction of the Gaussian measure is completed, from now on we will no longer distinguish between the spaces $\mathcal{H}$ and $\ell^{2}$):
\begin{equation}
    \int_\mathcal{H}e^{\frac{\varepsilon}{2}
    \Vert\varphi\Vert_{\mathcal{H}}^{2}}\,
    \mathcal{N}_{a,G}(d\varphi)=
    \begin{cases}{[\operatorname{Det}
    (1-\varepsilon\,G)]^{-1/2}
    e^{-\frac{\varepsilon}{2}
    \langle(1-\varepsilon\,G)^{-1}a|a\rangle},} 
    & \text { if } \varepsilon <\frac{1}{\lambda_1}, \\ +\infty, 
    & \text { otherwise. }\end{cases}   
\end{equation}

To calculate, we will again use the projector technique. For $\forall n \in \mathbb{N}$ we have:
\begin{equation}
    \int_\mathcal{H}e^{\frac{\varepsilon}{2}
    \Vert P_{n}\varphi\Vert_{\mathcal{H}}^{2}}\,
    \mathcal{N}_{a,G}(d\varphi)=
    \prod\limits_{k=1}^{n}\int_{\mathbb{R}}
    e^{\frac{\varepsilon}{2}\varphi_{k}^{2}} 
    N_{a_{k},\lambda_{k}}\left(d\varphi_{k}\right)=
    \prod\limits_{k=1}^{n}
    \frac{1}{\sqrt{1-\varepsilon\lambda_{k}}} 
    e^{-\frac{\varepsilon}{2}\frac{a_{k}^{2}}
    {1-\varepsilon\lambda_{k}}}.
\end{equation}

Since $\Vert P_{n}\varphi\Vert_{\mathcal{H}}^{2} \uparrow \Vert\varphi\Vert_{\mathcal{H}}^{2}$ for $n \rightarrow \infty $, the equality for the integral now follows from Beppo Levi’s monotone convergence
theorem (MCT).

Now let us consider the second example. Let $\psi \in \mathcal{H}$. Then the following equality holds:
\begin{equation}
    \int_\mathcal{H}e^{\langle\psi|\varphi\rangle}
    \mathcal{N}_{a,G}(d\varphi)=
    e^{\langle a|\psi\rangle+\frac{1}{2}\langle G\psi|\psi\rangle}    
\end{equation}

Indeed, for any $\varepsilon>0$ we have the following estimate:
\begin{equation}
    e^{\langle\psi|\varphi\rangle}\leq 
    e^{\Vert\psi\Vert_{\mathcal{H}}
    \Vert\varphi\Vert_{\mathcal{H}}}\leq 
    e^{\varepsilon\Vert\varphi\Vert_{\mathcal{H}}^{2}+
    \frac{1}{\varepsilon}\Vert\psi\Vert_{\mathcal{H}}^{2}} .    
\end{equation}

Let us choose $\varepsilon<\frac{1}{\lambda_1}$. By MCT 
we have the following chain of equalities:
\begin{equation}
\begin{split}
    &\int_\mathcal{H}e^{\langle\psi|\varphi\rangle}
    \mathcal{N}_{a,G}(d\varphi)=
    \lim\limits_{n\rightarrow\infty}\int_\mathcal{H}
    e^{\left\langle\psi|P_{n}\varphi\right\rangle}
    \mathcal{N}_{a,G}(d\varphi)=
    \lim\limits_{n\rightarrow\infty}\int_\mathcal{H} 
    e^{\left\langle\psi|P_{n}\varphi\right\rangle}
    \bigtimes\limits_{k=1}^{n} N_{a_{k},\lambda_{k}}(d\varphi)=\\
    &\lim\limits_{n\rightarrow\infty}
    e^{\left\langle P_{n}a|\psi\right\rangle+
    \frac{1}{2}\left\langle P_{n}G\psi|\psi\right\rangle}=
    e^{\left\langle a|\psi\right\rangle+
    \frac{1}{2}\langle G\psi|\psi\rangle}=\widehat{\mu}(i\psi).
\end{split}
\end{equation}

Thus, we learned to calculate the simplest integrals over the Gaussian measure in the separable Hilbert space $\mathcal{H}$. Let us note that such integrals coincide exactly with the integrals over $\ell^{2}$. This should be so, since the (invariant object) integral over the abstract $\mathcal{H}$ can be calculated after passing to a basis, the existence of which is guaranteed by the natural isomorphism $\gamma$. In this basis the integral is equal to that over $\ell^{2}$. The situation is similar, for example, to calculating the scalar product, which is specified axiomatically. But if a basis exists, it is equal to that in coordinates, therefore, it is calculated explicitly. The existence of the basis must be ensured independently.


\subsection{MCT and DCT}

Let a space with finite measure $(X,\mathscr{A},\mu)$ be given. Let us note that the theorems can be generalized to a wider class of measures, but this is not required in our paper. Let $(f_{n}(x))_{n=1}^{\infty}$ be a sequence of functions measurable with respect to $(X,\mathscr{A},\mu)$, that is $\forall n \in \mathbb{N}$ and $B \in \mathscr{B}\left(\mathbb{R}\right)$ it is true that $f_{n}^{-1}(B) \in \mathscr{A}$ (the inverse image of any Borel set is measurable). We will assume that all functions are defined on $A \in \mathscr{A}$ and that, in the general case, they can take infinite values.

Let us formulate MCT for non-negative functions. Let the inequality $0\leq f_{1}(x)\leq f_{2}(x)\leq\ldots\leq f_{n}(x)\leq\ldots$ be satisfied for any $x \in A$. Then for any $x \in A$ there is a limit function $f(x)=\lim\limits_{n\rightarrow\infty}f_{n}(x)$ (in the general case, it can take the value $+\infty $), and taking the integral and the limit can be interchanged (the value of integral equal to $+\infty$ is allowed):
\begin{equation}
  \int_{A}f(x)\mu(dx)=\lim\limits_{n\rightarrow\infty}
  \int_{A}f_{n}(x)\mu(dx).
\end{equation}

To go further, let us recall that a measure $\mu$ is said to be complete if $\forall A \in \mathscr{A}$ with measure $\mu(A)=0$ it is true that $\forall B \subset A$ measure $\mu(B)=0$.

Now let us formulate DCT. Let $\mu$ be a complete measure (the condition of the measure completeness can be replaced by the condition of the limit function $f(x)$ measurability; the condition of convergence almost everywhere can be replaced by the condition of convergence in measure), the sequence of functions $f_{n}(x) \rightarrow f(x)$ for $n\rightarrow\infty$ almost everywhere on $A$, and there exists a non-negative integrable (has finite integral value) function $F(x)$ such that for all $n \in \mathbb{N}$ and $x \in A$ $|f_{n}(x)|\leq F(x)$ (this function is called majorant). Then the limit function $f(x)$ is integrable (has finite integral value), and taking the integral and the limit can be interchanged:
\begin{equation}
  \int_{A}f(x)\mu(dx)=\lim\limits_{n\rightarrow\infty}
  \int_{A}f_{n}(x)\mu(dx).
\end{equation}

Using the above theorems, it is possible to carry out explicit calculations of integrals in infinite-dimensional spaces, reducing them to the limit of integrals over finite-dimensional subspaces, as was already done above.

\subsection{$\mathcal{S}$-Matrix Definition}

Now we are ready to specify the form of the integral over the Gaussian measure, which we will research further: $\mathcal{S}$-matrix of the theory. To do this, we introduce the interaction action of the $S$ theory in the form of integral of the Lagrangian with respect to the measure in the Euclidean $D$-dimensional space (the metric signature is all pluses): 
\begin{equation}
    S(g,\varphi)=g\langle\mathcal{L}(\varphi)\rangle=
    g\int_{\mathbb{R}^D}\mathbb{P}(dx)\mathcal{L}(\varphi(x)),
\end{equation}
where $\mathbb{P}$ is a probability measure, usually having density relative to the Lebesgue measure, $\mathcal{L}$ is the Lagrangian of the theory: even, strictly convex, continuous (one can increase the smoothness requirements if necessary), non-negative function with a single zero value at $\varphi=0$ and at $|\varphi|\rightarrow +\infty$ growing faster than $\varphi^{2}$. The angle brackets here mean averaging over the measure $\mathbb{P}$. In this case, the $\mathcal{S}$-matrix of the theory has the form (multiple Lebesgue integral with respect to variables $\left(\varphi_{a}\right)_{a=1}^{N}$, the $N$-fold integral, over $\mathbb{R}^{N}$ can be reduced to an iterated one using Fubini theorem, which explains the following notation):
\begin{equation}
    \mathcal{S}(g)=\int_{\mathcal{H}}\mathcal{N}_G(d\varphi)\, 
    e^{-g\langle\mathcal{L}(\varphi)\rangle}=
    \lim\limits_{N\rightarrow\infty}\left\{\prod\limits_{a=1}^{N}
    \int_{\mathbb{R}}\frac{d\varphi_a}{\sqrt{2\pi\lambda_a}}
    e^{-\frac{1}{2}\frac{\varphi_a^2}{\lambda_a}}\right\}
    e^{-g\int_{\mathbb{R}^D}\mathbb{P}(dx)
    \mathcal{L}\left(\sum\limits_{a=1}^{N}
    \varphi_a\psi_a(x)\right)}.
\end{equation}

The second equality is true by DCT. The system of functions $\psi_{a}$ is chosen in such a way that for the field $\varphi$ the pointwise (i.e. $\forall x \in \mathbb{R}^D$) equality $\varphi(x)=\sum\limits_{a=1}^{\infty}\varphi_a\psi_a(x)$ is true. This can always be achieved if we take some basis in $L^{2}(\mathbb{R}^D)$ with additional (damping) coefficients that make up an absolutely summable numerical sequence from $\ell^{1}$. Let us also note that summation over $a$ from $1$ to $\infty$ has a conventional notation: in the case $D>1$ this means summation over all components of the multi-index $a=\left(a_{1},\ldots,a_ {D}\right)$ within appropriate limits. But since this doesn't lead to a contradiction in our paper, in particular, it doesn't affect the applicability of DCT, we will use this convention.

In this paper we limit ourselves to the case of zero classical field -- the argument of the $\mathcal{S}$-matrix, which will not be indicated explicitly. However, we explicitly indicate the dependence of the $\mathcal{S}$-matrix on the coupling constant $g$.

Let us also note, that the coupling constant $g$ used in this paper is different from the coupling constant if the action $S$ is written in terms of the integral with respect to the Lebesgue measure. This is convenient for practical calculations, since in the latter case the presence of a ``box'' is often assumed, which allows one to introduce a probability measure. Another equivalent approach is to introduce a finite measure, the density of which relative to the Lebesgue measure is a ``smoothly switchable'' coupling constant. All these approaches make it possible to determine an action $S$ for which majorant estimates are applicable, which, in turn, is necessary to prove the existence of the $\mathcal{S}$-matrix of the theory.

\subsection{Linear Changes of Variables in Gaussian Integrals}

To complete the theoretical introduction, we consider what changes of variables are permissible in infinite-dimensional integrals with respect to the Gaussian measure. Let us study a linear diagonal change of variables that satisfies the condition that all diagonal elements are positive and their infinite product converges:
\begin{equation}
    \varphi_{a}=\frac{1}{\sqrt{c_{a}}}\chi_{a}, \quad c_{a}>0, 
    \quad \prod\limits_{a=1}^{\infty}c_{a}<+\infty.
    \label{change}
\end{equation}

Let us note that there is no summation over index $a$ in the right hand side of the first equality (\ref{change}). Let us also note that from the convergence of the infinite product it follows that $c_{a}\rightarrow 1$ for $a\rightarrow\infty$. Now we need to substitute the new variable into the integral and understand whether the limiting result can be interpreted as a new measure. Substitute the replacement into the integral:
\begin{equation}
    \mathcal{S}(g)=\lim\limits_{N\rightarrow\infty}
    \left\{\prod\limits_{a=1}^{N}
    \int_{\mathbb{R}}\frac{d\chi_a}{\sqrt{2\pi c_a\lambda_a}}\,
    e^{-\frac{1}{2}\frac{\chi_a^2}{\lambda_ac_a}}\right\}
    e^{-g\int_{\mathbb{R}^D}\mathbb{P}(dx)\mathcal{L}\left(\sum\limits_{a=1}^{N}
    \frac{\chi_a}{\sqrt{c_a}}\psi_a(x)\right)}.
\end{equation}

Suppose that $c_a\sim g^{k}\varepsilon_a$, where $k \in \mathbb{R}$. Next, we reduce the resulting integral to an integral over the original measure of such an integrand for which it is easy to explicitly highlight the asymptotics as $g\rightarrow\infty$, for example, as a power of the coupling constant $g$. At the moment, the factor $c_{a}$ in the measure is hindering us. Consider the transformation:
\begin{equation}
    \frac{1}{\sqrt{c_a}}e^{-\frac{\chi_a^2}{2\lambda_a}\frac{1}{c_a}}=
    e^{-\frac{1}{2}\ln{c_a}}e^{-\frac{\chi_a^2}{2\lambda_a}}
    e^{\frac{\chi_a^2}{2\lambda_a}(1-\frac{1}{c_a})}.
\end{equation}

After this transformation we return the original Gaussian measure:
\begin{equation}
    \mathcal{S}(g)=\lim\limits_{N\rightarrow\infty}
    \left\{\prod\limits_{a=1}^{N}
    \int_{\mathbb{R}}\frac{d\chi_a}{\sqrt{2\pi\lambda_a}}
    e^{-\frac{1}{2}\frac{\chi_a^2}{\lambda_a}}\right\}
    e^{-\frac{1}{2}\sum\limits_{a=1}^N\ln{c_a}}
    e^{\sum\limits_{a=1}^{N}\frac{\chi_a^2}{2\lambda_a}(1-\frac{1}{c_a})}
    e^{-g\int_{\mathbb{R}^D}\mathbb{P}(dx)
    \mathcal{L}\left(\sum\limits_{a=1}^{N}
    \frac{\chi_a}{\sqrt{c_a}}\psi_a(x)\right)}.
    \label{New111}
\end{equation}

We additionally require that the product $c_{a}$ converges to some predetermined number. Let us note that if the inequality $\frac{1}{c_a}>1$ holds, the mass term will be positive. Under all the assumptions made, the Hellinger integral $H$ will be positive:
\begin{equation}
    H=\prod\limits_{a=1}^{\infty}
    \frac{\sqrt{2\sqrt{c_{a}}}}{\sqrt{1+c_{a}}},
\end{equation}
which means that the initial and new Gaussian measures are equivalent. Since the measures are probabilistic, the largest possible value for $H$ is $H=1$. And although in our case the equivalence of the measures was verified directly, in the general case the Hellinger integral is a convenient mathematical tool, since in the infinite-dimensional case even such a harmless transformation as propagator change $G \rightarrow c\,G$ where $c>0$ and $c\neq 1$, generates nonequivalent measures (such measures are called singular).

Let us return to the derivation. Any sequence $c_{a}$ for which the equality $e^{-\frac{1}{2}\sum\limits_{a=1}^{\infty}\ln{c_a}}=\frac{ \alpha}{\sqrt[4]{g}}$ holds for some constant $\alpha$, satisfies the solution of the problem. After passing to the limit, the integral with respect to measure arises again:
\begin{equation}
    \lim\limits_{N\rightarrow\infty}
    \left\{\prod\limits_{a=1}^{N}
    \int_{\mathbb{R}}\frac{d\chi_a}{\sqrt{2\pi\lambda_a}}
    e^{-\frac{1}{2}\frac{\chi_a^2}{\lambda_a}}\right\}\ldots
    =\int_{\mathcal{H}}\mathcal{N}_G(d\chi)\ldots
\end{equation}

Since there are limits of the infinite product and integrals over finite measures, this allows us to write the limiting form of the expression (\ref{New111}) as follows (we take the factor depending on $g$ in a power-law manner out of the integral sign):
\begin{equation}
    \mathcal{S}(g)=
    e^{-\frac{1}{2}\sum\limits_{a=1}^{\infty}\ln{c_a}}
    \int_{\mathcal{H}}\mathcal{N}_G(d\chi)
    e^{-\frac{1}{2}\sum\limits_{a=1}^{\infty}
    \frac{\chi_a^2}{\lambda_a}(\frac{1}{c_a}-1)}
    e^{-g\int_{\mathbb{R}}\mathbb{P}(dx)
    \mathcal{L}\left(\sum\limits_{a=1}^{\infty}
    \frac{\chi_a}{\sqrt{c_a}}\psi_a(x)\right)}.
\end{equation}

Thus, we have highlighted the asymptotic behavior of the integral for $g\rightarrow +\infty$ (the asymptotic behavior of the integral can be easily found using the squeeze theorem). The rest is a function of $g$ that doesn't tend to zero at $g\rightarrow\infty$. Thus, using a linear substitution, we can extract the asymptotics from the integral. As for nonlinear substitutions, this issue requires a separate in-depth study.
\section{Perturbation Theory for the $\mathcal{S}$-Matrix and Bell Polynomials}
\label{sect:pert-theory}

\subsection{Perturbation Series}

We start with the following expression for the $\mathcal{S}$-matrix:
\begin{equation}
     \mathcal{S}(g)=\int_{\mathcal{H}}\mathcal{N}_G(d\varphi)\, 
     e^{-S(g,\varphi)}.
     \label{Z}
\end{equation}

Let us write the interaction action in general form:
\begin{equation}
    S(g,\varphi)=g\int_{\mathbb{R}^D}
    \mathbb{P}(dx)\mathcal{L}(\varphi(x)).
    \label{action}
\end{equation}

Next, let us expand the exponential function under the integral into the Taylor series, after which we interchange the integral and the sum, which will lead us to the asymptotic series of perturbation theory (PT):
\begin{equation}
    \mathcal{S}(g)=\sum\limits_{n=0}^{\infty}
    \frac{(-1)^{n}}{n!}\int_{\mathcal{H}}\mathcal{N}_G(d\varphi)
    S^n(g,\varphi)=\sum\limits_{n=0}^{\infty}
    \frac{(-1)^{n}}{n!}\,\mathcal{S}_{n}(g).
\end{equation}

In the last expression we introduced a notation $\mathcal{S}_{n}(g)$ for the integral, which is the Gaussian mean of the $n$th power of the interaction action $S$. Let us focus on researching this integral:
\begin{equation}
    \mathcal{S}_n(g)=\int_{\mathcal{H}}\mathcal{N}_G(d\varphi)
    S^n(g,\varphi).
\end{equation}

It is time to specify the type of interaction action so that we can move on, namely, choose the Lagrangian $\mathcal{L}(\varphi)$ in (\ref{action}). Let us make the following choice:
\begin{equation}
    \mathcal{L}(\varphi)=\left(2\,\sinh(\varphi)\right)^{2p},
    \quad p \in \mathbb{N}, \quad p>1.
\end{equation}

One could, as is customary, choose the Lagrangian $\varphi^{4}$. We chose the hyperbolic sine for a number of reasons. Firstly, PT series in such a model diverges as $e^{n^{2}}$, that is faster than any factorial power. Standard methods for asymptotic series summation are not applicable in this case. Secondly, the exponential representation of the hyperbolic sine is very convenient in combinatorial terms: It allows one to avoid the Isserlis--Wick combinatorics by writing the result for the PT $n$th order in terms of the corresponding Bell polynomial. The coefficient $2$ in front of the sine is added for the convenience of further calculations. Now we rearrange the $n$th power of the interaction action (we keep the interaction constant $g$ under the integral sign for ease of notation) as follows:
\begin{equation}
    S^n(g,\varphi)=\left\{\prod_{k=1}^{n}
    \int_{\mathbb{R}^D}g\cdot\mathbb{P}(dx_k)\right\}
    \left\{\prod_{k=1}^{n}\left(2\,\sinh
    (\varphi(x_k))\right)^{2p}\right\}.
\end{equation}

Multiple Lebesgue integral with respect to variables $\left(x_{k}\right)_{k=1}^{n}$, the $Dn$-fold integral, over $\mathbb{R}^{Dn}$ can be reduced to an iterated one using Fubini theorem (assume that $\mathbb{P}$ is a complete measure and the integrand is measurable with respect to the product measure $\mathbb{P}\times\mathbb{P}\times\cdots\times\mathbb{P}$ $n$ times), which explains the notation. Further, let us rewrite the degree of the hyperbolic sine in terms of a binomial:
\begin{equation}
    \left(2\,\sinh(\varphi)\right)^{2p}= 
    \left(e^\varphi-e^{-\varphi}\right)^{2p}=
    \sum\limits_{q=0}^{2p}C_{2p}^q(-1)^qe^{\varphi(2p-q)}
    e^{-\varphi q}=\sum\limits_{q=0}^{2p}C_{2p}^q(-1)^q
    e^{2(p-q)\varphi}.
\end{equation}

Thus, the following chain of equalities is valid:
\begin{equation}
\begin{split}
     &S^n(g,\varphi)=\left\{\prod_{k=1}^{n}\int_{\mathbb{R}^D}g\cdot\mathbb{P}(dx_k)\right\}\left\{\prod_{k=1}^{n}\sum\limits_{q=0}^{2p}C_{2p}^q(-1)^qe^{2(p-q)\varphi(x_k)}\right\}=\\
     &\left\{\prod_{k=1}^{n}\int_{\mathbb{R}^D}g\cdot\mathbb{P}(dx_k)\right\}\sum\limits_{q_1=0}^{2p}\ldots\sum\limits_{q_n=0}^{2p}(-1)^{q_1+\ldots +q_n}C_{2p}^{q_1}\ldots C_{2p}^{q_n}\,e^{2\sum\limits_{k=1}^{n}(p-q_k)\varphi(x_k)}.
\end{split}
\end{equation}

Consider the expansion of the field $\varphi$ over a basis in the space $L^{2}(\mathbb{R}^D)$ with additional (damping) coefficients that are needed to ensure pointwise convergence of the series:
\begin{equation}
    \varphi(x_k)=\sum\limits_{a=1}^{\infty}\varphi_a\psi_a(x_k).
\end{equation} 

In this expression, $\psi_a(x)$ are some basis functions, the exact form of which we can specify later. As a working example, it would be possible to use Hermite functions (wave functions of the quantum harmonic oscillator eigenstates) with damping coefficients. Let us note that the quantities $\varphi_a$ over which integration is performed don't depend on $x_k$, but they must be quadratically summable sequences, that is $(\varphi_a)_{a=1}^{\infty} \subset \ell^{2}$. Now, denoting $j_a(x)=\sum\limits_{k=1}^{n}(p-q_k)\psi_a(x_k)$, where $x=(x_k)_{k=1}^{n }$, we can rewrite the exponent:
\begin{equation}
    e^{2\sum\limits_{k=1}^{n}(p-q_k)\varphi(x_k)}=
    e^{2\sum\limits_{a=1}^{\infty}\varphi_a\sum\limits_{k=1}^{n}
    (p-q_k)\psi_a(x_k)}=e^{2\sum\limits_{a=1}^{\infty}
    \varphi_a j_a(x)}.
\end{equation}

From this expression it is clear that $j_a$ can be considered as a source. So, for $S^{n}$ we have the following expression:
\begin{equation}
    S^n(g,\varphi)=
    \left\{\prod_{k=1}^{n}
    \int_{\mathbb{R}^D}g\cdot\mathbb{P}(dx_k)\right\}
    \sum\limits_{q_1=0}^{2p}\ldots
    \sum\limits_{q_n=0}^{2p}(-1)^{q_1+\ldots +q_n}
    C_{2p}^{q_1}\ldots C_{2p}^{q_n}\,
    e^{\sum\limits_{a=1}^{\infty}\varphi_aj_a(x)}.
 \label{S^n}
\end{equation}

Then we can rewrite $\mathcal{S}_{n}$ as follows:
\begin{equation}
    \mathcal{S}_{n}(g)=
    \int_{\mathcal{H}}\mathcal{N}_G(d\varphi)
    \left\{\prod_{k=1}^{n}\int_{\mathbb{R}^D}
    g\cdot\mathbb{P}(dx_k)\right\}\sum\limits_{q_1=0}^{2p}\ldots
    \sum\limits_{q_n=0}^{2p}(-1)^{\sum\limits_{k=1}^{n}q_k}
    C_{2p}^{q_1}\ldots C_{2p}^{q_n}\,
    e^{\sum\limits_{a=1}^{\infty}\varphi_aj_a(x)}.
\end{equation}

Next, we can interchange the integrals using Fubini theorem and calculate the integral over the Gaussian measure $\mathcal{N}_G$ (this is the integral from the theoretical introduction):
\begin{equation}
    \int_{\mathcal{H}}\mathcal{N}_G(d\varphi)\,
    e^{\sum\limits_{a=1}^{\infty}\varphi_aj_a(x)}=
    e^{\frac{1}{2}\sum\limits_{a=1}^{\infty}
    \lambda_a\sum\limits_{k_1,k_2=1}^{n}
    4(q_{k_1}-p)(q_{k_2}-p)\psi_a(x_{k_1})\psi_a(x_{k_2})}.
\end{equation}

Finally, we can rewrite the expression for $\mathcal{S}_{n}$ in terms of only the $Dn$-fold integral:
\begin{equation}
    \mathcal{S}_{n}(g)=\left\{\prod_{k=1}^{n}
    \int_{\mathbb{R}^D}g\cdot\mathbb{P}(dx_k)\right\}
    \sum\limits_{q_1,\dots q_n=0}^{2p}
    w(q_1,\ldots q_n,p,n)\,
    e^{\sum\limits_{k_1,k_2=1}^{n}
    (q_{k_1}-p)(q_{k_2}-p)G(x_{k_1}, x_{k_2})}.
\end{equation}

In this expression we have used the notation for the propagator in coordinate notation (the $2$ coefficient is included for convenience)
\begin{equation}
    G(x_{k_1}, x_{k_2})=2\sum\limits_{a=1}^{\infty}
    \lambda_a\psi_a(x_{k_1})\psi_a(x_{k_2})
    \label{prop}
\end{equation}
and combinatorial coefficient
\begin{equation}
    w(q_1,\ldots q_n,p)=(-1)^{q_{1}+\ldots+ q_{n}}C_{2p}^{q_1}
    \ldots C_{2p}^{q_n}.
    \label{Comc_coef_fut}
\end{equation}

Let us shift the summation index $q_{i}\rightarrow q_{i}+p$ and denote the new combinatorial coefficient as follows:
\begin{equation}
    w'(q_1,\ldots q_n,p)=(-1)^{q_{1}+\ldots+q_{n}+np}C_{2p}^{q_1+p}
    \ldots C_{2p}^{q_n+p}.
\end{equation}

On this path we arrive at the following expression for $\mathcal{S}_{n}$:
\begin{equation}
\begin{split}
    \mathcal{S}_n(g)=\left\{\prod_{k=1}^{n}
    \int_{\mathbb{R}^D}g\cdot\mathbb{P}(dx_k)\right\}
    \sum\limits_{q_1,\dots q_n=-p}^{p}w'(q_1,\ldots q_n,p)\,
    e^{\sum\limits_{k_1,k_2=1}^{n}q_{k_1}q_{k_2}
    G(x_{k_1}, x_{k_2})}=\\
    =\sum\limits_{q_1,\dots q_n=-p}^{p}w'(q_1,\ldots q_n,p)g^{n}
    \int_{\mathbb{R}^D}\mathbb{P}(dx_1)\ldots
    \int_{\mathbb{R}^D}\mathbb{P}(dx_n)\,
    e^{\sum\limits_{k_1,k_2=1}^{n}q_{k_1}q_{k_2}
    G(x_{k_1}, x_{k_2})}.
\label{Zlfin}
\end{split}
\end{equation}

The resulting expression for $\mathcal{S}_{n}$ can be beautifully interpreted in terms of statistical physics. It represents a canonical partition function (CPF) with a fixed ``number of particles'' $n$. However, one should pay attention to the fact that the expression (quadratic form) in the exponent can change sign depending on the values of $q_{k}$. Thus, further in this section we will develop a statistical approach to research the resulting expression.

\subsection{Mayer Cluster Expansion}

For a fixed set $q=(q_{k})_{k=1}^{n}$, consider the $Dn$-fold integral, equivalently, the iterated integral that arose at the end of the previous subsection:
\begin{equation}
    \mathcal{S}_{n}(g,q)=g^{n}
    \int_{\mathbb{R}^D}\mathbb{P}(dx_1)\ldots
    \int_{\mathbb{R}^D}\mathbb{P}(dx_n)\,
    e^{\sum\limits_{k_1,k_2=1}^{n}q_{k_1}q_{k_2}
    G(x_{k_1},x_{k_2})}.
\end{equation}

Let us introduce the quantity $U_{k_1k_2} = q_{k_1}q_{k_2}G(x_{k_1},x_{k_2})$. Then $q_{k_i}$ can be interpreted as the charge of the $i$-th particle, and $x_i$ as its coordinate. In such a picture $U_{k_1k_2}=q_{k_1}q_{k_2}G(x_{k_1}, x_{k_2})$ corresponds to the interaction energy of particles with numbers $k_1$ and $k_2$. As a first approximation, we assume that the propagator $G(x_{k_1},x_{k_2})$, like $U_{k_1k_2}$, is translationally invariant, that is it depends only on the distance between the particles $x_{k_1}-x_{k_2}$. Because if in the definition of $G(x_{k_1},x_{k_2})$ we use Hermite functions damped by a power factor, then the Mehler function (Mehler propagator) arises. Mehler function also known from the problem of the quantum harmonic oscillator. If the power factor is close to unity, the propagator weakly depends on $x_{k_1}+x_{k_2}$, which is the reason for the translation invariant approximation. Let us extract the diagonal part from $U_{k_1k_2}$:
\begin{equation}
    e^{\sum\limits_{k_1,k_2=1}^{n}U_{k_1k_2}}=
    e^{\sum\limits_{k=1}^{n}U_{kk}}
    e^{2\sum\limits_{k_1<k_2}^{n}U_{k_1k_2}}=
    e^{\sum\limits_{k=1}^{n}q_k^2G(0)}
    e^{2\sum\limits_{k_1<k_2}^{n}U_{k_1k_2}}.
\label{diag}
\end{equation}

We use Mayer cluster expansion~\cite{Pathria,Zhou-Dai,Ruelle}:
\begin{equation}
\begin{split}
    e^{\sum\limits_{k_1,k_2=1}^{n}q_{k_1}q_{k_2}
    G(x_{k_1}, x_{k_2})}=
    e^{\sum\limits_{k=1}^{n}q_k^2G(0)}
    e^{2\sum\limits_{k_1<k_2}^{n}U_{k_1k_2}}=\\
    =e^{\sum\limits_{k=1}^{n}q_k^2G(0)}
    \prod\limits_{k_1<k_2}^{n}e^{2U_{k_1k_2}}=
    e^{\sum\limits_{k=1}^{n}q_k^2G(0)}
    \prod\limits_{k_1<k_2}^{n}(1+f_{k_1k_2}).
    \label{Mayer}
\end{split}    
\end{equation}

The commonly accepted notation $f_{k_1k_2}$ is the Mayer function, which is zero in the absence of interactions. Let us substitute it into $\mathcal{S}_{n}(g,q)$:
\begin{equation}
\begin{split}
    &\mathcal{S}_{n}(g,q)=g^{n}
    e^{\sum\limits_{k=1}^{n}q_k^2G(0)}
    \int_{\mathbb{R}^D}\mathbb{P}(dx_1)\ldots
    \int_{\mathbb{R}^D}\mathbb{P}(dx_n)\,
    \prod\limits_{k_1<k_2}^{n}(1+f_{k_1k_2})=
    g^{n}e^{\sum\limits_{k=1}^{n}q_k^2G(0)}\times\\
    &\int_{\mathbb{R}^D}\mathbb{P}(dx_1)\ldots
    \int_{\mathbb{R}^D}\mathbb{P}(dx_n)\,
    \left\{1+\sum\limits_{k_1<k_2}^{n}f_{k_1k_2}+
    \!\!\!\!\!\!\!\!\!\!\!\!\!\!
    \underset{k_1<k_3 \mbox{ \footnotesize \text{or} }
    k_1=k_3 \mbox{ \footnotesize \text{and} } k_2<k_4}
    {\sum\limits_{k_1<k_2}^{n}
    \sum\limits_{k_3<k_4}^{n}}
    \!\!\!\!\!\!\!\!\!\!
    f_{k_1k_2}f_{k_3k_4}+
    \ldots\right\}.
\label{grafs1}
\end{split}
\end{equation}


As can be seen from the first line in (\ref{grafs1}), the condition on the indices $k_1, k_2, k_3, k_4$ entangles the various sums. Such a condition can be extrapolated to the following terms, although, of course, the conditions on the indices will become more complicated. Each term in (\ref{grafs1}) can be associated with a graph with $n$ vertices. In such a representation, $f_{ij}$ connects the $i$ and $j$ vertices. For example, for $n=3$ we have:
\begin{equation}
\begin{split}
     &\mathcal{S}_{3}(g,q)=g^{3}e^{G(0)(q_1^2+q_2^2+q_3^2)}
     \int_{\mathbb{R}^D}\mathbb{P}(dx_1)
     \int_{\mathbb{R}^D}\mathbb{P}(dx_2)
     \int_{\mathbb{R}^D}\mathbb{P}(dx_3)\,
     \times\\
     &(1+f_{12}+f_{13}+f_{23}+
     f_{12}f_{13}+f_{12}f_{23}+
     f_{13}f_{23}+f_{12}f_{13}f_{23}).
     \label{n=3}
\end{split}
\end{equation}

The first term will then represent three isolated vertices:
\begin{equation}
    \int_{\mathbb{R}^D}\mathbb{P}(dx_1)
    \int_{\mathbb{R}^D}\mathbb{P}(dx_2)
    \int_{\mathbb{R}^D}\mathbb{P}(dx_3)\,
    \equiv \includegraphics[height=0.04\linewidth,valign=c]{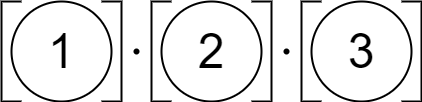}\, .
    \label{1.2.3.}
\end{equation}

The next three terms are $K_2$ (common notation for a complete graph with two vertices):
\begin{equation}
    \int_{\mathbb{R}^D}\mathbb{P}(dx_1)
    \int_{\mathbb{R}^D}\mathbb{P}(dx_2)
    \int_{\mathbb{R}^D}\mathbb{P}(dx_3)\,
    f_{12}\equiv
    \includegraphics[height=0.04\linewidth,valign=c]{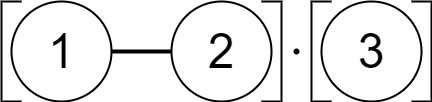}\, .
    \label{f12}
\end{equation}

The other three are triangular graphs missing one edge:
\begin{equation}
    \int_{\mathbb{R}^D}\mathbb{P}(dx_1)
    \int_{\mathbb{R}^D}\mathbb{P}(dx_2)
    \int_{\mathbb{R}^D}\mathbb{P}(dx_3)\,
    f_{12}f_{23}\equiv 
    \includegraphics[height=0.10\linewidth,valign=c]{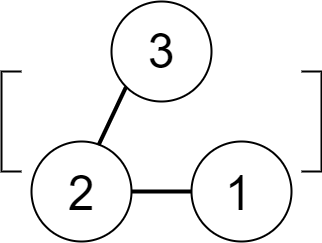}\, .
\end{equation}

The last term is $K_3$ (common notation for a complete graph with three vertices):
\begin{equation}
    \int_{\mathbb{R}^D}\mathbb{P}(dx_1)
    \int_{\mathbb{R}^D}\mathbb{P}(dx_2)
    \int_{\mathbb{R}^D}\mathbb{P}(dx_3)\,
    f_{12}f_{13}f_{23}\equiv 
    \includegraphics[height=0.10\linewidth,valign=c]{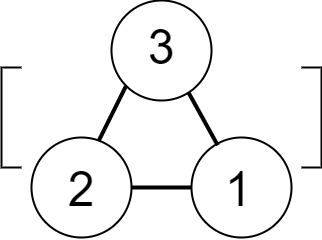}\, .
\end{equation}

In total, in the formula (\ref{grafs1}) there are $2^n$ terms under the integral, that is, the sum of $2^{n}$ graphs with $n$ vertices. Moreover, as can be seen from the examples above (\ref{1.2.3.}), (\ref{f12}), these graphs do not have to be connected.

Each term represents a configuration. For example, term (\ref{f12}) in the case $n=3$ is a configuration consisting of one 1-particle cluster and one 2-particle cluster. Obviously, as $n$ increases, the number of vertices will increase, and hence the number of clusters in the configuration. For example, for $n=8$, a configuration consisting of four 2-particle clusters has the following form:
\begin{equation}
\begin{split}
     &\int_{\mathbb{R}^D}\mathbb{P}(dx_1)\ldots
    \int_{\mathbb{R}^D}\mathbb{P}(dx_8)\,    
    f_{12}f_{34}f_{56}f_{78}\equiv\\
     &\includegraphics[height=0.04\linewidth,valign=c]{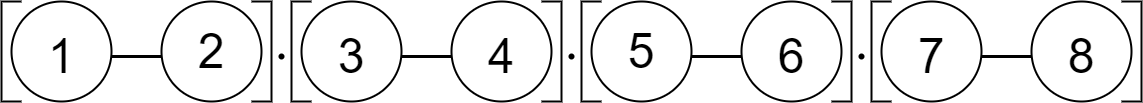}\, .
\end{split}
\end{equation}

Thus, an $n$-particle graph is a set of $n$ vertices with edges connecting them. Each edge is denoted by a pair of indices. Each term of the decomposition (\ref{grafs1}) is an $n$-vertex graph. Moreover, if we interchange the vertices in one term, we get a different graph, that is, this will be another term in the decomposition, which must also be taken into account. Now, since it is established that there is a one-to-one correspondence between the terms in the decomposition (\ref{grafs1}) and $n$-vertex graphs, we can say that $\mathcal{S}_{n}(g,q)=$ ``the sum of all distinct $n$-particle graphs''.

Next, we will factorize the various terms, for this we will introduce an $l$-cluster. An $l$-cluster is a connected graph with $l$ vertices. A cluster cannot be decomposed into simpler ones. In addition, $l$ particles can form different clusters, some of which can be equal in value. For example, in the case of $n=3$ (\ref{n=3}), the last $4$ terms are $3$-clusters, they are all different, but the values of the first three coincide. Given the variety of forms that an $l$-cluster can take, we introduce the definition of a cluster integral:
\begin{equation}
    b_l=\frac{1}{l!}\cdot (\text{sum of all possible $l$-clusters}).
    \label{cl_int}
\end{equation}

Here are the simplest cluster integrals:
\begin{equation}
    b_1=\includegraphics[height=0.04\linewidth,valign=c]{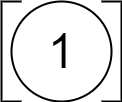}=\int_{\mathbb{R}^D}\mathbb{P}(dx_1)\equiv 1,
\end{equation}
\begin{equation}
    b_2=\frac{1}{2!}\includegraphics[height=0.04\linewidth,valign=c]{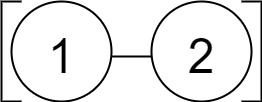}=\frac{1}{2}\int_{\mathbb{R}^D}\mathbb{P}(dx_1)\int_{\mathbb{R}^D}\mathbb{P}(dx_2)f_{12}.
\end{equation}

Using cluster integrals, we can rewrite (\ref{grafs1}). Again, using the example $n=3$ from (\ref{n=3}) for $\mathcal{S}_{3}(g,q)$ we have:
\begin{equation} 
    \mathcal{S}_{3}(g,q)=
    g^{3}e^{G(0)(q_1^2+q_2^2+q_3^2)}
    \left(1+b_2^{(12)}+b_2^{(13)}+b_2^{(23)}+b_3^{(123)}\right).
\end{equation}

Of course, an $n$-partial graph will consist of many clusters. Let $m_1$ be the number of $1$-clusters, $m_2$ be the number of $2$-clusters, $m_3$ be the number of $3$-clusters, and so on. In this case, the set $\{m_l\}$ must satisfy the restrictive condition:
\begin{equation}
    \sum\limits_{l=1}^nlm_l=n, \quad m_l\in \{0,\dots n\}.
    \label{mleq}
\end{equation}

The set $\{m_l\}$ does not determine the type of graph, it only determines the set of graphs, the total sum of which we denote by $S\{m_l\}$. Then we only need to calculate $\mathcal{S}_{n}(g,q)$ in terms of the sum of $S\{m_l\}$:
\begin{equation}
    \mathcal{S}_{n}(g,q)=
    g^{n}e^{G(0)(q_1^2+\ldots+q_n^2)}
    \sum\limits_{\{m_l\}}S\{m_l\}.
\end{equation}

This expression means the sum over all possible $\{m_l\}$ satisfying (\ref{mleq}). The difference between this expression and (\ref{grafs1}) is only in the rearrangement of the graphs. The next step is to calculate the sum $S\{m_l\}$. Let us note that there are many ways to form $\sum_l m_l$ clusters. To avoid taking one graph into account several times, we add a combinatorial factor:
\begin{equation}
    \frac{n!}{(1!)^{m_1}(2!)^{m_2}\dots(n!)^{m_n}}=
    \frac{n!}{\prod\limits_l (l!)^{m_l}}.
    \label{comb}
\end{equation}

Even for a given distribution of particles over clusters, a cluster of these particles can be formed in different ways. For example, the last $4$ terms in (\ref{n=3}) are $3$-clusters consisting of the same particles, but different ones. If all $l$-clusters were unique, then $S\{m_l\}$ would be calculated simply as the product of the combinatorial factor (\ref{comb}) and the value of the graphs:
\begin{equation}
    \prod\limits_l(\upsilon_l)^{m_l},
    \label{comb1}
\end{equation}
where $\upsilon_l$ is the value of the $l$-cluster. However, a correction is needed here due to the permutation of all particles of one cluster with all particles of another cluster of the same size, because this will be the same configuration. From these considerations, we add another combinatorial coefficient:
\begin{equation}
     \prod\limits_l \left(\frac{1}{m_l!}\right).
     \label{comb2}
\end{equation}

From the arguments above and the formulas (\ref{comb}), (\ref{comb1}) and (\ref{comb2}), let us assemble an expression for the sum of graphs $S\{m_l\}$:
\begin{equation}
    S\{m_l\}=n!\prod\limits_l^{n} \frac{(\upsilon_l)^{m_l}}{m_l!(l!)^{m_l}}.
\end{equation}

Thus, the sum of graphs can be rewritten explicitly using the cluster integral (\ref{cl_int}), which was introduced earlier:
\begin{equation}
    S\{m_l\}=n!\prod\limits_l^{n} \frac{(b_ll!)^{m_l}}{m_l!(l!)^{m_l}}=n!\prod\limits_l^{n}\frac{(b_l)^{m_l}}{m_l!}.
\end{equation}

Now we can write the result for the $Dn$-fold integral $\mathcal{S}_{n}(g,q)$:
\begin{equation}
    \mathcal{S}_{n}(g,q)=
    g^{n}e^{G(0)(q_1^2+\ldots+q_n^2)}\,
    n!\sum\limits_{\{m_l\}}
    \prod\limits_l^{n}\frac{(b_l)^{m_l}}{m_l!}=
    g^{n}e^{G(0)(q_1^2+\ldots+q_n^2)}
    B_{n}(X_1,\ldots,X_n).
\end{equation}

The last expression introduces the compact notation $X_l=l!b_l$, as well as $B_{n}$ for the $n$ (complete) Bell polynomial in $n$ variables which was already noted earlier in paper~\cite{Zhou-Dai}.

\subsection{Bell Polynomials and Bell Numbers}

The complete Bell polynomial $B_{n}$ is the following expression (we write $x_{i}$ instead of $X_{i}$ for convenience):
\begin{equation}
    B_{n}(x_1,\ldots,x_n)=n!\sum_{1j_{1}+\ldots+nj_{n}=n}
    \prod\limits_{i=1}^{n}\frac{x_{i}^{j_{i}}}{(i!)^{j_{i}}j_{i}!}.
\end{equation}

For ease of calculation, $B_{n}$ can be represented in terms of a determinant:
\begin{equation}
    B_{n}(x_1,\ldots,x_n)=\det
    \begin{bmatrix}
        \frac{x_1}{0!} & \frac{x_2}{1!} & \frac{x_3}{2!} & \frac{x_4}{3!} & \cdots & \cdots & \frac{x_n}{(n-1)!} \\  \\
        -1 & \frac{x_1}{0!} & \frac{x_2}{1!} & \frac{x_3}{2!} &  \cdots & \cdots & \frac{x_{n-1}}{(n-2)!} \\  \\
        0 & -2 & \frac{x_1}{0!} & \frac{x_2}{1!} & \cdots & \cdots & \frac{x_{n-2}}{(n-3)!} \\  \\
        0 & 0 & -3 & \frac{x_1}{0!} & \cdots  & \cdots & \frac{x_{n-3}}{(n-4)!} \\  \\
        0 & 0 & 0 & -4 & \cdots & \cdots & \frac{x_{n-4}}{(n-5)!} \\  \\
        \vdots & \vdots & \vdots &  \vdots & \ddots & \ddots & \vdots  \\  \\
        0 & 0 & 0 & 0 & \cdots & -(n-1) & \frac{x_1}{0!} 
    \end{bmatrix}.
\end{equation}

Here are some examples of the first first polynomials $B_{n}$:
\begin{equation}
    \begin{split}
        & B_{0} = 1, \,\,\,\, B_{1}(x_1) = x_1, \,\,\,\, 
        B_{2}(x_1,x_2) = x_1^2 + x_2, \,\,\,\, 
        B_{3}(x_1,x_2,x_3) = x_1^3 + 3x_1 x_2 + x_3, \\
        & B_{4}(x_1,x_2,x_3,x_4) = x_1^4 + 6 x_1^2 x_2 + 
        4 x_1 x_3 + 3 x_2^2 + x_4.
    \end{split}
\end{equation}

Now let us introduce the Bell numbers (the first equality is the definition):
\begin{equation}
    B_{n}=B_{n}(1,\ldots,1)=\frac{1}{e}\sum_{j=0}^\infty \frac{j^{n}}{j!}.
\end{equation}

For Bell numbers there is a simple majorant:
\begin{equation}
    B_{n} < \left( \frac{0.792 n}{\ln( n+1)} \right)^{n},\quad 
    n \in \mathbb{N}.
\end{equation}

The following asymptotic behavior of Bell numbers in terms of the Lambert function $W$ is also true (the growth rate of $W$ is the same as that of the logarithm) for $n\rightarrow\infty$:
\begin{equation}
    B_{n}\sim \frac{1}{\sqrt{n}} \left( \frac{n}{W(n)} \right)^{n + \frac{1}{2}} \exp\left(\frac{n}{W(n)} - n - 1\right). 
\end{equation}

Bell numbers can be used to determine the number of cluster integrals, therefore, to estimate the terms of the PT series. The resulting divergence is of order $e^{n^{2}}$. Thus, in order to obtain legitimate mathematical results, it is absolutely necessary to construct another series (as it turns out, in powers of interaction action $S$), which is what the next (main) section of our paper is devoted to.
\section{$\mathcal{S}$-Matrix in Terms of Iterated Series}
\label{sect:scattering-matrix}

\subsection{First Series Expansion}

Our main task is to calculate the integral over a Gaussian measure with a nuclear covariance operator (with non-degenerate variance) over a separable HS. The Gaussian measure is probabilistic, that is $\mathcal{N}_G(\mathcal{H})=1$.

Various approaches have been developed to calculate such integrals. For example, using the Fourier transform and the Parseval -- Plancherel equality, one can obtain the integral of an infinite product of functions that change sign depending on the value of their arguments (alternating sign functions). Such an integral is much more difficult to calculate than the original one. However, the Fourier transform is well suited for estimating the integral. We will show that in the theories considered in this paper this estimate is $\sim g^{-1/2p}$ for $p \in \mathbb{N}$ and $g\rightarrow+\infty$.

It is also possible to construct a PT series, for example, for polynomial theories $\varphi^{2p}$ for $p \in \mathbb{N}$, whose $n$th term grows as $n!$ (in some power depending on $p$), that is, no convergence is out of the question. In such cases, it is customary to sum asymptotic series. However, the appearance of an asymptotic series indicates the illegality of transformations in the calculations. Indeed, the original integral exists and positive. This is easy to show using majorant and minorant condition. This contradicts the PT result.

Primarily, we consider the theory $\sinh^4(\varphi)$, since it is easier to calculate Gaussian means of $S^{n}$, and the Lagrangian grows faster than $\varphi^{4}$. For such a theory, the divergence of the perturbation theory series is stronger -- $e^{n^2}$. We don't have a Borel-type summation method for such series~\cite{efimov1970nonlocal}, so we will construct some other series in terms of the interaction action degrees. Apparently, a multiple series of this type doesn't exist, since the latter must be a PT series, which, in turn, diverges.

The starting point of the derivation is again the expression for the $\mathcal{S}$-matrix of the theory:
\begin{equation}
     \mathcal{S}(g)=\int_{\mathcal{H}}
     \mathcal{N}_G(d\varphi)\, 
     e^{-g\langle\mathcal{L}(\varphi)\rangle}.
     \label{Z2}
\end{equation}

Recall that $\langle\mathcal{L}(\varphi)\rangle$ is the probabilistic measure  $\mathbb{P}$ mean:
\begin{equation}
   \langle\mathcal{L}(\varphi)\rangle=
   \int_{\mathbb{R}^D}\mathbb{P}(dx)
   \mathcal{L}\left(\varphi(x)\right), \quad
   \varphi(x)=\sum\limits_{a=1}^{\infty}
   \varphi_{a}\psi_{a}(x).
   \label{lagr}
\end{equation}

The functions $\psi_{a}$ are a damped basis in $L^{2}(\mathbb{R}^D)$, that is the series in (\ref{lagr}) converges pointwise ($\forall x \in \mathbb{R}^D$). The simplest thing we can do with the integrand in (\ref{Z2}) is to multiply it by a ``smart one'':
\begin{equation}
     \mathcal{S}(g)=\int_{\mathcal{H}}\mathcal{N}_G(d\varphi)\, 
     e^{-g\langle\mathcal{L}(\varphi)\rangle}\frac{R_1(\varphi)}{R_1(\varphi)}.
\end{equation}

Moreover, a ``smart one'' can depend on anything. What do we want from this multiplier? First, of course, it shouldn't be zero. Secondly, it is convenient to consider the integrand sign-constant and positive. In this case, we rewrite equivalently $R_1(\varphi)=e^{-\varepsilon_1(\varphi)}$, after which we obtain:
\begin{equation}
     \mathcal{S}(g)=
     \int_{\mathcal{H}}\mathcal{N}_G(d\varphi)\,
     e^{\varepsilon_1(\varphi)}
     e^{-\left[g\langle\mathcal{L}(\varphi)\rangle+
     \varepsilon_1(\varphi)\right]}.
     \label{Ze}
\end{equation}

We made the identity transformation. For example, $\varepsilon_1$ can be considered as a polynomial, since the exponential of a polynomial is measurable and integrable with respect to a Gaussian measure. The exponent to a negative power decreases rapidly, it ``controls'' the convergence of the integral. Next, let us introduce the notation:
\begin{equation}
     z(\varphi):=g\langle\mathcal{L}(\varphi)\rangle+\varepsilon_1(\varphi)=
     \frac{1}{(w(z(\varphi)))^\alpha} \quad \Rightarrow \quad 
     w(z)=z^{-\frac{1}{\alpha}}.
     \label{w}
\end{equation}

Essentially, the sum on the left hand side of the equation is a new action that we will work with. Since there is a risk of getting divergence of $w$ for small $\varphi$, we require that our new action be separated from zero. Therefore, we assume that $\varepsilon_1(\varphi)>0$ for $\forall \varphi \in \mathbb{R}$, in particular, $\varepsilon_1(0)>0$. As a trial run, we also require that $\varepsilon_1$ be bounded from above. This leads to the simplest choice for $\varepsilon_1$ being a constant.

Thus, we apply the experience of calculating Feynman diagrams, but now not in integrals over $\mathbb{R}^D$, but directly in the integral over $\mathcal{H}$: to get a convenient series expansion, we need to multiply/divide by something or add/subtract something. In particular, for numerical calculation it may be convenient to introduce the parameter $\alpha>0$.

Next, we expand the exponential from (\ref{Ze}) in some (we will define the explicit form later) family of polynomials $P_{m}$, having as their argument the function $w$:
\begin{equation}
    f(w(z(\varphi))):=e^{-\frac{1}{(w(z(\varphi)))^\alpha}}=
    \sum\limits_{m=0}^{\infty}a_m(\alpha)P_m(w(z(\varphi))).
    \label{series1}
\end{equation}

We represent the polynomials $P_{m}$ in the form:
\begin{equation}
    P_m(w)=\sum\limits_{l=0}^{m}u_l(m)w^{l}.
    \label{poli1}
\end{equation}

Why did we choose polynomial expansion rather than power expansion? For example, it is known that the power functions system $\left\{1, x, x^2, \dots\right\}$ is complete, but is not an orthonormal basis in $L^{2}[-1,1]$. The Gram -- Schmidt orthogonalization process transforms the power functions system into orthogonal Legendre polynomials, which are already a basis in $L^{2}[-1,1]$. Now we substitute the series (\ref{series1}) into the integral and interchange the summation and integration:
\begin{equation}
     \mathcal{S}(g)=
     \sum\limits_{m=0}^{\infty}a_m(\alpha)
     \int_{\mathcal{H}}\mathcal{N}_G(d\varphi)\,
     e^{\varepsilon_1(\varphi)}P_m(w(z(\varphi))).   
     \label{Zpol1}
\end{equation}

The main question is why, unlike PT, is it legal to interchange summation and integration? Before answering this question, let us substitute the polynomial (\ref{poli1}) and the explicit form of the function $w(z)=z^{-\frac{1}{\alpha}}$ into this expression:
\begin{equation}
\begin{split}
        &\mathcal{S}(g)=
        \sum\limits_{m=0}^{\infty}a_m(\alpha)
        \int_{\mathcal{H}}\mathcal{N}_G(d\varphi)\,
        e^{\varepsilon_1(\varphi)}P_m
        \left(\left[g\langle\mathcal{L}(\varphi)\rangle+
        \varepsilon_1(\varphi)\right]^{-\frac{1}{\alpha}}\right)=\\
        &\sum\limits_{m=0}^{\infty}a_m(\alpha)
        \sum\limits_{l=0}^{m}u_l(m)
        \int_{\mathcal{H}}\mathcal{N}_G(d\varphi)\,
        e^{\varepsilon_1(\varphi)}
        \left[g\langle\mathcal{L}(\varphi)\rangle+
        \varepsilon_1(\varphi)\right]^{-\frac{l}{\alpha}}.
        \label{Zser}
\end{split}
\end{equation}

Obviously, we can always interchange a finite sum with an integral, but with the infinite sum in (\ref{Zpol1}) things are not so obvious. The argument of the polynomial $\left[g\langle\mathcal{L}(\varphi)\rangle+\varepsilon_1(\varphi)\right]^{-\frac{1}{\alpha}}$ varies from $0$ to $(\varepsilon_1(0))^{-\frac{1}{\alpha}}$. The function $e^{-\frac{1}{w^{\alpha}}}$, satisfying the condition of continuous differentiability ($1$-smoothness), is expanded in terms of polynomials on a compact set, therefore, such an expansion converges uniformly. 

Let us note that if the image of the function $w(z)=z^{-\frac{1}{\alpha}}$ were not compact, but represented the entire $\mathbb{R^{+}}$, the convergence of the expansion in polynomials of the function $e^{-\frac{1}{w^{\alpha}}}$ would be pointwise, and therefore it would be impossible to interchange the summation and integration (in the general case). In our case, the series converges uniformly, which means that we can interchange the sum and the integral with respect to the Gaussian measure. Indeed, from the uniform convergence of the series (\ref{series1}) on the set $W=\left[0,(\varepsilon_1(0))^{-\frac{1}{\alpha}}\right]$ follows uniform convergence on $\mathcal{H}$ such that $w:\mathcal{H}\rightarrow W$. This is a simple generalization of the corresponding statement from mathematical analysis about functions on $\mathbb{R}$. But then one can find the following estimate of the integrals $\forall\delta>0$:
\begin{equation}
\begin{split}
    &\bigg|\int_{\mathcal{H}}\mathcal{N}_G(d\varphi)\,f(w(z(\varphi)))-
    \int_{\mathcal{H}}\mathcal{N}_G(d\varphi)
    \sum\limits_{m=0}^{n}a_m(\alpha)P_m(w(z(\varphi)))\bigg| \leq \\
    &\int_{\mathcal{H}}\mathcal{N}_G(d\varphi)
    \bigg|f(w(z(\varphi)))-\sum\limits_{m=0}^{n}a_m(\alpha)
    P_m(w(z(\varphi)))\bigg| \leq \delta.
\end{split}
\end{equation}

Thus, the interchange of summation and integration is justified. Further, we will consider $\varepsilon_1(\varphi)=\varepsilon(g)$ independent of $\varphi$, but we will emphasize its dependence on $g$. The purpose of the function $\varepsilon(g)$ is to ensure separability from zero.

\subsection{Second Series Expansion}

Let us do the trick by multiplying the integrand in (\ref {Zser}) by ``smart one'' again, which we have already done earlier:
\begin{equation}
     \mathcal{S}(g)=
     \sum\limits_{m=0}^{\infty}a_m(\alpha)
     \sum\limits_{l=0}^{m}u_l(m)
     \frac{1}{g^{\frac{l}{\alpha}}}
     \int_{\mathcal{H}}\mathcal{N}_G(d\varphi)
     \frac{e^{\varepsilon(g)}}{\left[\langle\mathcal{L}(\varphi)\rangle+
     \frac{1}{g}\,\varepsilon(g)\right]^\frac{l}{\alpha}}
     \frac{R_2(\varphi)}{R_2(\varphi)}.
\end{equation}

The function $R_2(\varphi)$ in the general case can also depend on anything, but it is convenient to choose it so that the argument of the polynomial is separated from infinity (without violating separability from zero). In this case, it will again be possible to prove the validity of the interchange of the sum and the integral. It turns out to be convenient to choose $R_2(\varphi)=e^{-\frac{l}{\alpha}\varepsilon_2(\varphi,g)}$. The specific form of $\varepsilon_2(\varphi,g )$ will be chosen later. Next, we consider the integral, which we denote as $\mathcal{I}(g)$:
\begin{equation}
    \begin{split}
    &\mathcal{I}(g)=\int_{\mathcal{H}}\mathcal{N}_G(d\varphi)
    \frac{e^{\varepsilon(g)}}{\left[\langle\mathcal{L}(\varphi)\rangle+
    \frac{1}{g}\,\varepsilon(g)\right]^\frac{l}{\alpha}}
    \frac{R_2(\varphi)}{R_2(\varphi)}=\\
    &e^{\varepsilon(g)}\int_{\mathcal{H}}\mathcal{N}_G(d\varphi)
    \frac{e^{-\frac{l}{\alpha}\varepsilon_2(\varphi,g)}}
    {\left\{\left[\langle\mathcal{L}(\varphi)\rangle+\frac{1}{g}\,
    \varepsilon(g)\right]
    e^{-\varepsilon_2(\varphi,g)}\right\}^\frac{l}{\alpha}}.
    \end{split}
\end{equation}

We introduce a notation for the new combination in the denominator of the integral:
\begin{equation}
    \left[\langle\mathcal{L}(\varphi)\rangle+\frac{1}{g}\,
    \varepsilon(g)\right]e^{-\varepsilon_2(\varphi,g)}=\omega(\varphi).
    \label{ser2}
\end{equation}

Let us again perform an expansion (this time of the denominator) by some (we will define the explicit form later) family of polynomials $Q_{i}$, having as their argument the function $\omega$:
\begin{equation}
    \frac{1}{(\omega(\varphi))^\frac{l}{\alpha}}=\sum\limits_{i=0}^{\infty}b_i\left(\frac{l}{\alpha}\right)Q_i(\omega(\varphi)).
\end{equation}

We represent the polynomials $Q_{i}$ in the form:
\begin{equation}
    Q_i(\omega)=\sum\limits_{j=0}^{i}v_q(i)\omega^{j}.
\end{equation}

Let us rewrite $\mathcal{I}(g)$ through a polynomial expansion, interchanging the sum and integral in the first line:
\begin{equation}
\begin{split}
    &\mathcal{I}(g)=e^{\varepsilon(g)}\sum\limits_{i=0}^{\infty}
    b_i\left(\frac{l}{\alpha}\right)\int_{\mathcal{H}}\mathcal{N}_G(d\varphi)\,
    e^{-\frac{l}{\alpha}\varepsilon_2(\varphi,g)}
    Q_i\left(\left[\langle\mathcal{L}(\varphi)\rangle+
    \frac{1}{g}\,\varepsilon(g)\right]
    e^{-\varepsilon_2(\varphi,g)}\right)=\\
    &e^{\varepsilon(g)}\sum\limits_{i=0}^{\infty}
    b_i\left(\frac{l}{\alpha}\right)\sum\limits_{j=0}^{i}v_j(i)
    \int_{\mathcal{H}}\mathcal{N}_G(d\varphi)\,
    e^{-\frac{l}{\alpha}\varepsilon_2(\varphi,g)}
    \left[\langle\mathcal{L}(\varphi)\rangle+\frac{1}{g}\,
    \varepsilon(g)\right]^j
    e^{-j\varepsilon_2(\varphi,g)}=\\
    &e^{\varepsilon(g)}\sum\limits_{i=0}^{\infty}
    b_i\left(\frac{l}{\alpha}\right)\sum\limits_{j=0}^{i}v_j(i)
    \int_{\mathcal{H}}\mathcal{N}_G(d\varphi)\,
    e^{-(\frac{l}{\alpha}+j)\varepsilon_2(\varphi,g)}
    \sum\limits_{n=0}^{j}C_{j}^{n}
    \langle\mathcal{L}(\varphi)\rangle^n\left(\frac{1}{g}\,
    \varepsilon(g)\right)^{j-n}.
    \label{SerToInt}
\end{split}
\end{equation}

And again we ask ourselves: Why can we interchange the sum and the integral in the first line (\ref{SerToInt})? This time we use a corollary of MCT for series. That is, the transition is legitimate if the following series of Gaussian means converges:
\begin{equation}
    \sum\limits_{i=0}^{\infty}\left|b_i\left(\frac{l}{\alpha}\right)\right|
    \int_{\mathcal{H}}\mathcal{N}_G(d\varphi)\,
    e^{-\frac{l}{\alpha}\varepsilon_2(\varphi,g)}
    |Q_i(\omega(\varphi))|<+\infty.
\end{equation}

The convergence of this series for a properly chosen family of polynomials $Q_{i}$ will be shown below. Substituting now the expression (\ref{SerToInt}) for $\mathcal{I}(g)$ into the original $\mathcal{S}$-matrix, we obtain an iterated series over the pair of indices $(m,i)$. The algorithm with the introduction of an integrating factor could be repeated further, and one could even suggest that a wide class of integrals with respect to a Gaussian measure with a nuclear covariance operator over a separable HS can be calculated in a finite number of such steps. However, the integral we have obtained in two iterations is already computable. At this moment, we have obtained a result in terms of a series in powers of interaction action. A similar result also arises in PT, but with different parameters, which leads to its divergence.

Now we need to choose $\varepsilon_2(\varphi,g)$ so that this integral can be calculated. Strictly speaking, when choosing $\varepsilon_2(\varphi,g)$, we need to take into account two conditions: so that we can calculate the integral in the last line of the expression (\ref{SerToInt}), and so that it is legal  to interchange the integral and the sum in the first line in the same expression. For example, the Gaussian exponent satisfies these requirements. One way to look at it is that we have ``renormalized'' the vacuum energy and the mass term (the Gaussian exponent). But this ``renormalization'' turns out to be an exact equality, a ``non-perturbative renormalization''.

\subsection{Asymptotics for $i\rightarrow\infty$ and Proof of Interchange}

As a family of polynomials $Q_{i}$ we choose the generalized Laguerre polynomials:
\begin{equation}
    Q_i(\omega):= L_i^{(\gamma)}(\omega).
\end{equation}

These are orthogonal polynomials in the space $L^{2}(\mathbb{R},\omega^\gamma e^{-\omega}d\omega)$. In such a space, the scalar product is defined as follows:
\begin{equation}
    \forall f,g \in L^{2}(\mathbb{R},\omega^\gamma e^{-\omega}d\omega)
    \quad \langle f|g\rangle=\int_{0}^{+\infty}
    d\omega\,\omega^\gamma e^{-\omega}f(\omega)g(\omega).
\end{equation}

Since we are expanding the function $\omega^{-\frac{l}{\alpha}}$ (for a fixed $l \in \mathbb{N}$) in generalized Laguerre polynomials, it is necessary that $\omega^{-\frac{l}{\alpha}} \in L^{2}(\mathbb{R},\omega^\gamma e^{-\omega}d\omega)$, which implies that
\begin{equation}
    \int_{0}^{+\infty}d\omega \,\omega^{\gamma-2\frac{l}{\alpha}} 
    e^{-\omega}<+\infty.  
\end{equation}

For this integral to converge, it is necessary that $\gamma > 2\frac{l}{\alpha}-1$. This gives the first constraint on $\gamma$. Let us note that as the value of $l$ increases, the value of $\gamma$ must also increase, that is different functions $\omega^{-\frac{l}{\alpha}}$ are expanded into different families of $L_i^{(\gamma)}$.

Further, if the function $f \in L^{2}(\mathbb{R},\omega^\gamma e^{-\omega}d\omega)$, then the expansion in generalized Laguerre polynomials converges to it in the norm of this space. Under additional conditions on $f$, one can achieve pointwise convergence of the expansion in generalized Laguerre polynomials to $f$. Such conditions are set out, for example, in Theorem $9.1.5.$ from G. Szego's book ``Orthogonal Polynomials''~\cite{Szego}. Let us formulate the corresponding theorem.

Let $f$ be a Lebesgue measurable function on $[0,+\infty)$ and let there exist Lebesgue integrals
\begin{equation}
    \int_{0}^{1}d\omega\,\omega^\gamma |f(\omega)| \mbox{ and }
    \int_{0}^{1}d\omega\,\omega^{\frac{\gamma}{2}-
    \frac{1}{4}} |f(\omega)|.
\end{equation}

Let it also be true that
\begin{equation}
    \int_{n}^{+\infty}d\omega\, e^{-\frac{\omega}{2}}
    \omega^{\frac{\gamma}{2}-
    \frac{13}{12}}|f(\omega)|=o(n^{-\frac{1}{2}}),
    \quad n\rightarrow\infty.
\end{equation}

Then, if $s_{n}(f,\omega)$ denotes the $n$th partial sum of the expansion in generalized Laguerre polynomials for $f$, then for $\omega>0$ the following equality holds:
\begin{equation}
    \lim\limits_{n\rightarrow\infty}
    \left\{s_{n}(f,\omega)-\frac{1}{\pi}
    \int_{\sqrt{\omega}-\delta}^{\sqrt{\omega}+\delta}d\tau
    f(\tau^{2})\,\frac{\sin{[2\sqrt{n}(\sqrt{\omega}-\tau)]}}
    {\sqrt{\omega}-\tau}\right\}=0.
\end{equation}

Here $\delta$ is a fixed positive number such that $\delta<\sqrt{\omega}$. Moreover, the last relation is fulfilled uniformly on the positive segment $\delta_{1}\leq\omega\leq\delta_{2}$, where $\delta<\sqrt{\delta_{1}}$. Let us note that the book also contains a variation of this theorem, but it will not be needed in our paper.

Thus, for the function $f(\omega)=\omega^{-\frac{l}{\alpha}}$ (for a fixed $l \in \mathbb{N}$) one can achieve pointwise convergence of its expansion in a series of generalized Laguerre polynomials. To do this, it is sufficient to require the already mentioned condition $\gamma > 2\frac{l}{\alpha}-1$.

Next, we want to apply a corollary of MCT for series. This requires that the function $f$ be from $L^{1}(\mathbb{R},\omega^\gamma e^{-\omega}d\omega)$. Given the condition on $\gamma$, this is ensured automatically, since the norm of the function $f$ in the space $L^{1}(\mathbb{R},\omega^\gamma e^{-\omega}d\omega)$ turns out to be finite.

Now we find the coefficients of the expansion of $f$ in terms of generalized Laguerre polynomials (the slope $\varGamma$ denotes the gamma function):
\begin{equation}
    b_i\left(\frac{l}{\alpha}\right)=
    \frac{\left\langle \omega^{-\frac{l}{\alpha}}\big|
    L_i^{(\gamma)}(\omega)\right\rangle}
    {\left\langle L_i^{(\gamma)}(\omega)\big|
    L_i^{(\gamma)}(\omega)\right\rangle}=
    \frac{\varGamma(i+1)}{\varGamma(i+\gamma+1)}
    \int_{0}^{+\infty}d\omega\,
    e^{-\omega}\omega^{\gamma-\frac{l}{\alpha}}
    L_i^{(\gamma)}(\omega).
    \label{bi}
\end{equation}

The generalized Laguerre polynomial of degree $i$, the explicit form of which can be obtained from Rodrigues' formula by differentiating according to Leibniz rule, reads:
\begin{equation}
    L_i^{(\gamma)}(\omega)=
    \sum\limits_{s=0}^{i}\frac{(-1)^s}{s!}\,
    C_{i+\gamma}^{i-s}\,\omega^{s}.
\end{equation}

Let us substitute the polynomial in this form into the coefficient $b_i\left(\frac{l}{\alpha}\right)$:
\begin{equation}
     b_i\left(\frac{l}{\alpha}\right)=
     \frac{\varGamma(i+1)}{\varGamma(i+\gamma+1)}
     \sum\limits_{s=0}^{i}\frac{(-1)^s}{s!}\,
     C_{i+\gamma}^{i-s}\,\int_{0}^{+\infty}d\omega\, 
     e^{-\omega}\omega^{\gamma+s-\frac{l}{\alpha}}.
\end{equation}

The integral in this expression is easy to calculate -- it is the gamma function $\varGamma$:
\begin{equation}
    \int_{0}^{+\infty}d\omega\, 
    e^{-\omega}\omega^{\gamma+s-\frac{l}{\alpha}}=
    \varGamma\left(\gamma-\frac{l}{\alpha}+s+1\right).
\end{equation}

Substituting this gamma function into the expression (\ref{bi}) and taking the final sum over $s$, we obtain the coefficient in the following form:
\begin{equation}
     b_i\left(\frac{l}{\alpha}\right)=
     \frac{\varGamma(i+\frac{l}{\alpha})
     \varGamma(1-\frac{l}{\alpha}+\gamma)}
     {\varGamma(\frac{l}{\alpha})\varGamma(i+\gamma+1)}
     \sim \frac{\varGamma(1-\frac{l}{\alpha}+\gamma)}
     {\varGamma(\frac{l}{\alpha})}i^{\frac{l}{\alpha}-\gamma-1}, 
     \quad i\rightarrow\infty. 
\end{equation}

The obtained asymptotics will be sufficient to check the convergence of the Gaussian means series, since the latter is of constant sign.

Now let us find an upper bound for generalized Laguerre polynomials. First of all, we can choose $\varepsilon_2(\varphi,g)$ such that $\omega(\varphi)\rightarrow 0$ as $\varphi_{a}\rightarrow\infty$ for $\forall a \in \mathbb{N}$, reaching its maximum $\mathcal{M}$ at some point $\varphi^{0}=(\varphi_{a}^{0})_{a=1}^{\infty}$. This maximum doesn't depend on $i$, $j$ and other parameters, it can depend only on $g$. Let us return again to G. Szego's book ``Orthogonal Polynomials''~\cite{Szego} and now use formula $7.6.9$ for $i\rightarrow\infty$, and the upper (in book) equality, since the lower equality turns out to be too rough an estimate, which doesn't allow us to move further:
\begin{equation}
    \forall i \in \mathbb{N} \quad 
    \forall \omega\in (0,\mathcal{M}] \quad
    |L_i^{(\gamma)}(\omega)|
    \leq\frac{B_\gamma(g)}{\omega^{\frac{\gamma}{2}+
    \frac{1}{4}}}\, p^{\frac{\gamma}{2}-\frac{1}{4}}.
    \label{leqSz}
\end{equation}

Here $B_\gamma(g)$ is some constant of $\varphi$, which also doesn't depend on $i$. We substitute the obtained asymptotics for $b_i\left(\frac{l}{\alpha}\right)$ and the estimate from (\ref{leqSz}) into the series of Gaussian means, which is a majorant for the integral $\mathcal{I}(g)$:
\begin{equation}
    |\mathcal{I}(g)|\sim
    \sum\limits_{i=1}^{\infty}
    \frac{\varGamma(1-\frac{l}{\alpha}+\gamma)}
    {\varGamma(\frac{l}{\alpha})}\,
    i^{(\frac{l}{\alpha}-\frac{\gamma}{2}-\frac{5}{4})}
    \int_{\mathcal{H}}\mathcal{N}_G(d\varphi)
    B_{\gamma}(g)\frac{e^{\left(\frac{\gamma}{2}+\frac{1}{4}-
    \frac{l}{\alpha}\right)\varepsilon_2(\varphi,g)}}
    {\left[\langle\mathcal{L}(\varphi)\rangle+\frac{1}{g}\,
    \varepsilon(g)\right]^{\frac{\gamma}{2}+\frac{1}{4}}}.
    \label{Isim}
\end{equation}

Let us consider separately the series and the integral over the Gaussian measure. The series in $i$ converges if the following condition on $\gamma$ is satisfied:
\begin{equation}
    -\frac{l}{\alpha}+\frac{\gamma}{2}+\frac{5}{4}>1 
    \quad \Rightarrow \quad 
    \gamma > 2\frac{l}{\alpha}-\frac{1}{2}.
    \label{gamma}
\end{equation}

Let us note that this is the most stringent condition on $\gamma$, so we will focus on it. In this case, the series in (\ref{Isim}) converges to the Riemann zeta function:
\begin{equation}
    \zeta\left(-\frac{l}{\alpha}+\frac{\gamma}{2}+
    \frac{5}{4}\right)=
    \sum\limits_{i=1}^{\infty}\frac{1}
    {i^{\left(-\frac{l}{\alpha}+\frac{\gamma}{2}+\frac{5}{4}\right)}}, 
    \quad  \gamma > 2\frac{l}{\alpha}-\frac{1}{2}.
\end{equation}

Now it only remains to consider the integral over the Gaussian measure in expression (\ref{Isim}). From the condition on $\gamma$ in expression (\ref{gamma}) we see that the factor in front of $\varepsilon_2$ in the exponential is positive, we also chose $\varepsilon_2$ to be positive to obtain a Gaussian exponential, and $\omega(\varphi)\rightarrow 0$ for $\varphi_{a}\rightarrow\infty$ for $\forall a \in \mathbb{N}$. At first glance, it seems that this integral diverges. It's time to recall the properties of the Gaussian measure.

Let us choose a specific type of $\varepsilon_2$. What limitations do we have? If this function grows faster than $\varphi^{2}$, that is, the integrand is ``stronger'' than the Gaussian exponent, the integral will diverge. Therefore, we take the following form for $\varepsilon_2$:
\begin{equation}
    \varepsilon_2(\varphi,g)=\eta(g)+m^{2} 
    \Vert \varphi \Vert_{\ell^{2}}^{2}, \quad
    \Vert \varphi \Vert_{\ell^{2}}^{2}=
    \sum\limits_{a=1}^{\infty}\varphi_{a}^2.
    \label{e2}
\end{equation}

Here $\eta(g)$ doesn't depend on $\varphi$. Thus, we again limited ourselves to the simplest, that polynomial in $\varphi$ dependence. Further, the integral of interest to us in the expression (\ref{Isim}) is the integral of a non-negative function. Such an integral always exists in $\overline{\mathbb{R}}$ (extended real number system). Let us write a majorant for this integral, simply decreasing the denominator (as a trial run, we can assume $\eta(g)=0$):
\begin{equation}
\begin{split}
    &\int_{\mathcal{H}}\mathcal{N}_G(d\varphi)\,
    \frac{e^{\left(\frac{\gamma}{2}+\frac{1}{4}-
    \frac{l}{\alpha}\right)\varepsilon_2(\varphi,g)}}
    {\left[\langle\mathcal{L}(\varphi)\rangle+\frac{1}{g}\,
    \varepsilon(g)\right]^{\frac{\gamma}{2}+\frac{1}{4}}}\leq \int_{\mathcal{H}}\mathcal{N}_G(d\varphi)\,
    \frac{e^{\left(\frac{\gamma}{2}+\frac{1}{4}-
    \frac{l}{\alpha}\right)\varepsilon_2(\varphi,g)}}
    {\left(\frac{1}{g}\,
    \varepsilon(g)\right)^{\frac{\gamma}{2}+\frac{1}{4}}}=\\
    &\left(\frac{g}{\varepsilon(g)}\right)^{\frac{\gamma}{2}+
    \frac{1}{4}}e^{\left(\frac{\gamma}{2}+\frac{1}{4}-
    \frac{l}{\alpha}\right)\eta(g)}
    \int_{\mathcal{H}}\mathcal{N}_G(d\varphi)\,
    e^{\left(\frac{\gamma}{2}+\frac{1}{4}-
    \frac{l}{\alpha}\right)m^2
    \Vert \varphi \Vert_{\ell^{2}}^{2}}.
\end{split}
\end{equation}

We have already calculated this last integral in the theoretical introduction using MCT:
\begin{equation}
    \int_{\mathcal{H}}\mathcal{N}_G(d\varphi)\,
    e^{\left(\frac{\gamma}{2}+\frac{1}{4}-
    \frac{l}{\alpha}\right)m^2
    \Vert \varphi \Vert_{\ell^{2}}^{2}}=
    \lim\limits_{N\rightarrow\infty}
    \left\{\prod\limits_{a=1}^{N}
    \int_{\mathbb{R}}\frac{d\varphi_a}{\sqrt{2\pi\lambda_a}}\,
    e^{-\left(\frac{1}{2\lambda_a}-
    (\frac{\gamma}{2}+\frac{1}{4}-
    \frac{l}{\alpha})m^2\right)\varphi_a^2}\right\}.
    \label{int_lim}
\end{equation}

For this integral to converge, the following condition must be satisfied:
\begin{equation}
    \frac{1}{2\lambda_a}-\left(\frac{\gamma}{2}+
    \frac{1}{4}-\frac{l}{\alpha}\right)m^2 > 0.
    \label{Cond1}
\end{equation}

Since the eigenvalues of $G$ are ordered in a non-increasing manner:
\begin{equation}
    \lambda_1 \geq \lambda_2 \geq \cdots \geq \lambda_n \geq \cdots > 0 
    \quad \Rightarrow \quad 0 < \frac{1}{2\lambda_1} \leq \frac{1}{2\lambda_2} \leq \cdots \leq \frac{1}{2\lambda_n} \leq \cdots
    \label{lamb_upor}
\end{equation}

From expressions (\ref{Cond1}) and (\ref{lamb_upor}) we obtain the following condition on the parameter $m$:
\begin{equation}
    0<\left(\frac{\gamma}{2}+\frac{1}{4}-
    \frac{l}{\alpha}\right)m^2 < \frac{1}{2\lambda_1}.
\end{equation}

From this condition we obtain an upper bound for $m$:
\begin{equation}
    m^2 < m_c^2 = \frac{\frac{1}{\lambda_1}}
    {\gamma+\frac{1}{2}-2\frac{l}{\alpha}}.
\end{equation}

The value $m_c$ can naturally be called the ``critical mass''. We see that if the ``mass'' $m$ is less than some ``critical mass'', then such an integral (\ref{int_lim}) turns out to be finite, which allows us to present a majorant for the integral $\mathcal{I}(g)$. Let us note that the chosen method of proof uses a sufficient condition for the interchange of summation and integration. There may be other estimates for $m_c$ or its analogues. In any case, the legitimacy of the interchange is fully proven and convergent iterated series is obtained.

\subsection{$\mathcal{S}$-Matrix in Terms of Iterated Series and Bell Polynomials}

Let us return to the formula (\ref{SerToInt}) for $\mathcal{I}(g)$, namely, to its last line:
\begin{equation}
    e^{\varepsilon(g)}
    \sum\limits_{i=0}^{\infty}b_i\left(\frac{l}{\alpha}\right)
    \sum\limits_{j=0}^{i}v_j(i)
    \sum\limits_{n=0}^{j}C_{j}^{n}
    \left(\frac{1}{g}\,\varepsilon(g)\right)^{j-n}
    \!\!\int_{\mathcal{H}}\mathcal{N}_G(d\varphi)\,
    e^{-\left(\frac{l}{\alpha}+j\right)\varepsilon_2(\varphi,g)}
    \langle\mathcal{L}(\varphi)\rangle^{n}.
    \label{SerToInt2}
\end{equation}

In the second part of the paper, almost the same integral was calculated in terms of Bell polynomials. In this expression there is an additional factor $e^{-(\frac{l}{\alpha}+j)\varepsilon_2(\varphi,g)}$ under the integral sign. The function $\varepsilon_2(\varphi,g)$ is chosen as in expression (\ref{e2}). We substitute it into expression (\ref{SerToInt2}), after which we denote the integral as $\mathcal{J}(g)$. Next we consider this integral:
\begin{equation}
    \mathcal{J}(g)=\int_{\mathcal{H}}\mathcal{N}_G(d\varphi)\,
    e^{-\left(\frac{l}{\alpha}+j\right)
    \Vert \varphi \Vert_{\ell^{2}}^{2}}
    \langle\mathcal{L}(\varphi)\rangle^{n}.
    \label{I2}
\end{equation}

We choose the Lagrangian $\sinh^{2p}$, where $p \in \mathbb{N}$. The powers of interaction action with such a Lagrangian are given in expression (\ref{S^n}). In the formula (\ref{I2}) $\langle\mathcal{L}(\varphi)\rangle=\frac{S(g,\varphi)}{g}$, so that
\begin{equation}
\begin{split}
    &\mathcal{J}(g)=\int_{\mathcal{H}}\mathcal{N}_G(d\varphi)\,
    e^{-\left(\frac{l}{\alpha}+j\right)\Vert \varphi \Vert_{\ell^{2}}^{2}}
    \left\{\prod_{k=1}^{n}\int_{\mathbb{R}^D}\mathbb{P}(dx_k)\right\}
    \sum\limits_{q_1\ldots q_n =0}^{2p}w(q_1,\ldots, q_n,p)\,
    e^{\sum\limits_{a=1}^{\infty}\varphi_aj_a(x_k)}\\
    &=\left\{\prod_{k=1}^{n}\int_{\mathbb{R}^D}\mathbb{P}(dx_k)\right\}
    \sum\limits_{q_1\ldots q_n =0}^{2p}w(q_1,\ldots, q_n,p)
    \int_{\mathcal{H}}\mathcal{N}_G(d\varphi)\,
    e^{-\left(\frac{l}{\alpha}+j\right)\Vert \varphi \Vert_{\ell^{2}}^{2}}
    e^{\sum\limits_{a=1}^{\infty}\varphi_aj_a(x_k)}.
\end{split}    
\end{equation}

Here we again use the combinatorial coefficient $w(q_1,\ldots, q_n,p)$ from the expression (\ref{Comc_coef_fut}). We again received a Gaussian integral, which can be calculated explicitly:
\begin{equation}
    \int_{\mathcal{H}}\mathcal{N}_G(d\varphi)\,
    e^{-\left(\frac{l}{\alpha}+j\right)\Vert \varphi \Vert_{\ell^{2}}^{2}}
    e^{\sum\limits_{a=1}^{\infty}\varphi_aj_a(x_k)}=
    \left\{\prod\limits_{a=1}^{\infty}\frac{1}{\sqrt{1+2\lambda_a\left(\frac{l}{\alpha}+j\right)}}\right\}
    e^{\frac{1}{2}\sum\limits_{a=1}^{\infty}
    \frac{j_a^2(x_k)\lambda_a}
    {1+2\lambda_a\left(\frac{l}{\alpha}+j\right)}}.
\label{I22}
\end{equation}

Recall that the source $j_a(x)=\sum\limits_{k=1}^n(p-q_k)\psi_a(x_k)$, where $x=(x_k)_{k=1}^{n}$. Let us substitute this into (\ref{I22}) and shift all indices $q_{i}$ by $p$:
\begin{equation}
      \mathcal{J}(g)=\left\{\prod_{k=1}^{n}
      \int_{\mathbb{R}^D}\mathbb{P}(dx_k)\right\}
      \sum\limits_{q_1\ldots q_n =-p}^{p}w'(q_1,\ldots, q_n,p)
      \left\{\prod\limits_{a=1}^{\infty}
      \frac{1}{\sqrt{1+2\lambda_a
      \left(\frac{l}{\alpha}+j\right)}}\right\}
      e^{\mathcal{Q}_{n}(x)},
\end{equation}
where compact notations are introduced
\begin{equation}
      w'(q_1,\ldots, q_n,p)=(-1)^{q_1+\ldots +q_n+np}
      C_{2p}^{q_1+p}\ldots C_{2p}^{q_n+p}
\end{equation}
for the new combinatorial coefficient and
\begin{equation}
    G\left(x_{k_1},x_{k_2};\frac{l}{\alpha}+j\right)=
    2\sum\limits_{a}^{\infty}
    \frac{\lambda_a\psi_a(x_{k_1})\psi_a(x_{k_2})}
    {1+2\lambda_a\left(\frac{l}{\alpha}+j\right)}, \quad
    \mathcal{Q}_{n}(x)=\sum\limits_{k_1,k_2=1}^{n}
    q_{k_1}q_{k_2}G\left(x_{k_1},x_{k_2};\frac{l}{\alpha}+j\right)
\end{equation}
for a new propagator containing a ``polarization operator'' in the denominator, suppressing propagator for large $l$ and $j$, which turns out to be essential for the convergence of the series, and the corresponding quadratic form.

Further calculations are absolutely analogous to the calculations in the section~\ref{sect:pert-theory} of the paper: as in the expression (\ref{diag}), we separate the diagonal part, and then, as in the expression (\ref{Mayer}), we rewrite the exponent through the Mayer functions. After which we obtain the Bell polynomials. We immediately write down the result for $\mathcal{J}(g)$:
\begin{equation}
    \mathcal{J}(g)=\sum\limits_{q_1\ldots q_n =-p}^{p}
    w'(q_1,\ldots, q_n,p)\left\{\prod\limits_{a=1}^{\infty}
    \frac{1}{\sqrt{1+2\lambda_a\left(\frac{l}
    {\alpha}+j\right)}}\right\}
    e^{\mathcal{Q}_{n}}B_n(X_1,\ldots,X_n),
\end{equation}
where compact notations are introduced
\begin{equation}
    \mathcal{Q}_{n}=
    \sum\limits_{k=1}^{n}q_k^{2}\,
    G\left(x_{k},x_{k};\frac{l}{\alpha}+j\right)=
    \sum\limits_{k=1}^{n}q_k^{2}\,
    G\left(0;\frac{l}{\alpha}+j\right)
\end{equation}
for the diagonal part of the quadratic form $\mathcal{Q}_{n}$ in the translation invariant approximation, but taking into account the polarization operator, and
\begin{equation}
    X_i=i!b_i=\sum\limits_{\Gamma\in\mathbb{G}_{c,i}}
    \int_{\mathbb{R}^D}\mathbb{P}(dx_{1})\ldots
    \int_{\mathbb{R}^D}\mathbb{P}(dx_{i})
    \prod\limits_{k_1<k_2\in E(\Gamma)}
    f_{k_1k_2}^{\nu_{k_1k_2}(\Gamma)}
\end{equation}
for the argument of the Bell polynomial. The last expression requires some explanation: the sum $\sum\limits_{\Gamma\in\mathbb{G}_{c,i}}$ is computed over all connected graphs with $i$ vertices, $\nu_{k_1k_2}(\Gamma)$ is the adjacency matrix of the graph, equal to one if there is an edge between $k_1$ and $k_2$, and zero otherwise, $\mathbb{G}_{c,i}$ is the set of all connected graphs with $i$ vertices, $E(\Gamma)$ is the set of all vertices (the vertices are numbered) in the graph, $\Gamma$ is the notation of the graph (we use the upright $\Gamma$ to distinguish the notation of the graph from the gamma function $\varGamma$), $b_i$ is the cluster integral, which was defined in the expression (\ref{cl_int}).

So, putting it all together, we get the final result for the $\mathcal{S}$-matrix in terms of the iterated series and Bell polynomials. This expression is not asymptotic, which means that what follows is a matter of numerical calculations and approximations from statistical physics. For example, one can use the approximations of hard-sphere gas and stochastic clusters, and Bell numbers to estimate the number of cluster integrals.

\subsection{Majorant and Minorant for $\mathcal{S}$-Matrix for large $g$}

To get a complete picture for the $\mathcal{S}$-matrix, we need to evaluate its behavior for large values of $g$. Let the Lagrangian of the theory satisfy the following requirements: $\mathcal{L}(\varphi)$ is an even, strictly convex, continuous (smoothness conditions can be added along the way), non-negative function with a single zero value at $\varphi=0$. Let us find an estimate for the mean over the measure $\mathbb{P}$ of the value $\langle\mathcal{L}(\varphi)\rangle$:
\begin{equation}
    \langle\mathcal{L}(\varphi)\rangle=
    \int\limits_{\mathbb{R}^D}\mathbb{P}(dx)
    \mathcal{L}(\varphi(x)).
\end{equation}

Since the Lagrangian is an even field function, the following is true:
\begin{equation}
    \mathcal{L}(\varphi(x))=\mathcal{L}(|\varphi(x)|)=
    \mathcal{L}\left(|\sum\limits_{a=1}^{\infty}
    \varphi_a\psi_a(x)|\right)\leq
    \mathcal{L}\left(\sum\limits_{a=1}^{\infty}
    |\varphi_a||\psi_a(x)|\right),
\end{equation}
where the functions $\psi_{a}$ satisfy the inequality $|\psi_a(x)|\leq\mu_a$ for $\forall a \in \mathbb{N}$, that is each of them is bounded. So, using H\"{o}lder inequality, we obtain a simple estimate:
\begin{equation}
    \langle\mathcal{L}(\varphi(x))\rangle\leq\mathcal{L}
    \left(\sum\limits_{a=1}^{\infty}|\varphi_a|\mu_a\right)\leq
    \mathcal{L}\left(\Vert\varphi\Vert_p\Vert\mu\Vert_q\right), 
    \quad p\geq 1, \quad \frac{1}{p}+\frac{1}{q}=1.
\end{equation}

Let the sequence $\mu=(\mu_{a})_{a=1}^{\infty} \in \ell^{q}$. In this case, its norm will be denoted by $\Vert\mu\Vert_q\equiv\mu<+\infty$. Thus, the following estimate is true:
\begin{equation}
    \langle\mathcal{L}(\varphi(x))\rangle\leq
    \mathcal{L}\left(\mu\left\{\sum\limits_{a=1}^{\infty}
    |\varphi_a|^p\right\}^{\frac{1}{p}}\right)=
    \mathcal{L}\left(\sqrt[p]{\mu^{p}
    \sum\limits_{a=1}^{\infty}|\varphi_a|^p}\right).
\end{equation}

Let us introduce the compact notation $t=\mu^{p}\sum\limits_{a} |\varphi_a|^p$ and write the minorant of the integrand:
\begin{equation}
    e^{-g\langle\mathcal{L}(\varphi)\rangle}\geq 
    e^{-g\mathcal{L}(\sqrt[p]{|t|})}=
    \int_{\mathbb{R}}\frac{d\xi}{2\pi}\,
    F_{p,g}(\xi)\cos{(t\xi)}.
\end{equation}

The last equality contains the Fourier transform for the function $e^{-g\mathcal{L}(\sqrt[p]{|t|})}$. Such a transformation is an ordinary (not a distribution) function if the minorant is Lebesgue integrable with respect to the variable $t$. Further we consider this condition to be fulfilled. Next, we write down the minorant for the $\mathcal{S}$-matrix:
\begin{equation}
   \int_{\mathcal{H}}\mathcal{N}_G(d\varphi)\,
   e^{-g\langle\mathcal{L}(\varphi)\rangle}\geq 
   \int_{\mathcal{H}}\mathcal{N}_G(d\varphi)
   \int_{\mathbb{R}}\frac{d\xi}{2\pi}\,
   F_{p,g}(\xi)\cos{(t\xi)}.
\end{equation}

According to Fubini theorem, we interchange the integrals over $\xi$ and the Gaussian measure $\mathcal{N}_{G}$. Given the value of $p$, the integral over $\mathcal{N}_{G}$ could be calculated, but for now we limit ourselves to introducing the notation:
\begin{equation}    
    A_{p}(\xi):=\int_{\mathcal{H}}\mathcal{N}_G(d\varphi)
    \cos{(t\xi)}.
\label{Ap}
\end{equation}

Therefore, the following estimate is true:
\begin{equation}
    \int_{\mathcal{H}}\mathcal{N}_G(d\varphi)e^{-g\langle\mathcal{L}(\varphi)\rangle}\geq\int_{\mathbb{R}}\frac{d\xi}{2\pi}\,
    F_{p,g}(\xi)A_p(\xi).
    \label{minor}
\end{equation}

Let us write explicitly $F_{p,g}(\xi)$ for the theory with the Lagrangian $\varphi^4$:
\begin{equation}
    F_{p,g}(\xi)=\int_{\mathbb{R}}dy\,
    e^{-i\xi y-g|y|^{\frac{4}{p}}}=
    \int_{\mathbb{R}}dy\,
    e^{-g|y|^{\frac{4}{p}}}\cos{(\xi y)}.
\end{equation}

Let us make the substitution $y=g^{\alpha}t$, choosing $\alpha=-\frac{p}{4}$. In this case, the dependence on $g$ in the function $F_{p,g}(\xi)$ is expressed explicitly (the last equality is the definition of the function $\Phi_{p}$):
\begin{equation}
    F_{p,g}(\xi)=\frac{1}{g^{\frac{p}{4}}}
    \int_{\mathbb{R}}dt\, e^{-|t|^{\frac{4}{p}}}
    \cos{\left(\frac{\xi t}{g^{\frac{p}{4}}}\right)}=
    \frac{1}{g^{\frac{p}{4}}}
    \Phi_p\left(\frac{\xi}{g^{\frac{p}{4}}}\right).
    \label{Phip}
\end{equation}

Thus, we arrive at the following estimate:
\begin{equation}
    \int_{\mathcal{H}}\mathcal{N}_G(d\varphi)\,
    e^{-g\langle\mathcal{L}(\varphi)\rangle}\geq 
    \frac{1}{g^{\frac{p}{4}}}
    \int_{\mathbb{R}}\frac{d\xi}{2\pi}\,
    \Phi_p\left(\frac{\xi}{g^{\frac{p}{4}}}\right)A_p(\xi).
\end{equation}

Let us choose $p=1$ and estimate the functions $\Phi_{1}$ and $A_{1}$. For the function $A_{1}$, using the explicit expression for $t$, rewriting the integral over $\mathcal{N}_{G}$ through the limit and calculating the latter, we obtain the following expression:
\begin{equation}
    A_1(\xi)=\left\{\prod\limits_{a=1}^{\infty}
    e^{-\frac{1}{2}\lambda_a (\mu\xi)^2}
    \sqrt{1+\operatorname{Erfi}^2
    \left(\frac{\sqrt{\lambda_a}\mu\xi}{\sqrt{2}}\right)}\right\}
    \cos{\left(\sum\limits_{a=1}^{\infty}
    \arctan{\operatorname{Erfi}\left(\frac{\sqrt{\lambda_a}\mu\xi}{\sqrt{2}}\right)}\right)}.
\end{equation}

Let us find the majorant of such an expression. The cosine is bounded by one, and the infinite product is bounded by a finite $\forall N \in \mathbb{N}$:
\begin{equation}
    |A_1(\xi)|<e^{N\frac{1}{2}\lambda_N\mu\xi}
    \left(1+\operatorname{Erfi}^2\left(\frac{\sqrt{\lambda_N}\mu\xi}{\sqrt{2}}\right)\right)^{\frac{N}{2}}\sim
    \left(\frac{\sqrt{2}}{\sqrt{\pi\lambda_N}\mu\xi}\right)^N\sim
    \frac{1}{\xi^N}, \quad \xi\rightarrow\infty.
\end{equation}

As we see, the function $A_1$ decreases faster than any power, that is, $A_1\in L^1(\mathbb{R})$. Next, let us consider $\Phi_1$:
\begin{equation}
    \Phi_1(\eta)=\int_{\mathbb{R}}dt\, e^{-t^4}
    \cos{(\eta t)}=\Phi_1+\Delta\Phi_1(\eta),
\end{equation}
where is the additional function
\begin{equation}
    \Delta\Phi_1(\eta)=-\int_{\mathbb{R}}dt\, e^{-t^4}
    2\sin^2{\left(\frac{\eta t}{2}\right)}=O(\eta^2), 
    \quad \eta\rightarrow 0. 
\end{equation}

We arrive at the following estimate:
\begin{equation}
    \frac{1}{g^{\frac{1}{4}}}
    \int_{\mathbb{R}}\frac{d\xi}{2\pi}\,
    \Phi_1\left(\frac{\xi}{g^{\frac{1}{4}}}\right)
    A_1(\xi)\sim\frac{1}{\sqrt[4]{g}}
    \frac{\varGamma\left(\frac{5}{4}\right)}{\pi}
    \int_{\mathbb{R}}d\xi A_1(\xi), 
    \quad g\rightarrow+\infty. 
\end{equation}

This entails the following estimate for the $\mathcal{S}$-matrix:
\begin{equation}
    \int_{\mathcal{H}}\mathcal{N}_G(d\varphi)\,
    e^{-g\langle\mathcal{L}(\varphi)\rangle}\geq 
    \frac{1}{\sqrt[4]{g}}
    \frac{\varGamma\left(\frac{5}{4}\right)}{\pi}
    \int_{\mathbb{R}}d\xi A_1(\xi).
    \label{Phi-4_min}
\end{equation}

Thus, we have obtained a lower bound $\sim g^{-1/4}$ for the $\mathcal{S}$-matrix of the theory with Lagrangian $\varphi^4$. Now consider a theory with an arbitrary Lagrangian, adding the condition $\forall\varphi\in\mathbb{R}$ Lagrangian $\mathcal{L}(\varphi)\geq\varphi^{2(r+1)}$ with some $r\in\mathbb{N}$. In this case, the following is true:
\begin{equation}
    \langle\mathcal{L}(\varphi)\rangle \geq \langle\varphi^{2(r+1)}\rangle 
    \Rightarrow e^{-g\langle\mathcal{L}(\varphi)\rangle} \leq 
    e^{-g\langle\varphi^{2(r+1)}\rangle},
\end{equation}
what does inequality entail
\begin{equation}
    \int_{\mathcal{H}}\mathcal{N}_G(d\varphi)\,
    e^{-g\langle\mathcal{L}(\varphi)\rangle}\leq
    \int_{\mathcal{H}}\mathcal{N}_G(d\varphi)\,
    e^{-g\langle\varphi^{2(r+1)}\rangle}.
\end{equation}

This means that the majorant of the theory $\sinh^4(\varphi)$ is exactly the majorant of the theory $\varphi^4$. The majorant of the latter, which can be shown again using the Fourier transform, is estimated to be $\sim g^{-1/4}$. Next we find the minorant similarly to the theory $\varphi^4$ (\ref{minor}). Let $\mathcal{L}(\sqrt[p]{|y|})=\mathcal{L}(|y|)=\sinh^4{y}$, which corresponds to the choice $p=1$. Then the Fourier transform has the form:
\begin{equation}
    F_{1,g}(\xi)=\int_{\mathbb{R}}dy\,
    e^{-i\xi y}e^{-g\sinh^4y}.
\end{equation}

By changing variables, this expression is transformed to the form:
\begin{equation}
     F_{1,g}(\xi)=\frac{1}{\sqrt[4]{g}}
     \int_{\mathbb{R}}\frac{d\eta}
     {\sqrt{1+\frac{1}{\sqrt{g}}}\eta}\,
     e^{-\eta^4}\cos{\left(\frac{\xi}{2}\,\operatorname{arsh}
     {\frac{\eta}{\sqrt[4]{g}}}\right)}\sim
     \frac{1}{\sqrt[4]{g}}\int_{\mathbb{R}}d\eta\, 
     e^{-\eta^4}.
\end{equation}

This in turn means the validity of the following expression:
\begin{equation}
    \int_{\mathcal{H}}\mathcal{N}_G(d\varphi)\,
    e^{-g\langle\mathcal{L}(\varphi)\rangle}\geq
    \int_{\mathbb{R}}\frac{d\xi}{2\pi}\,
    F_{1,g}(\xi)A_1(\xi)\sim
    \frac{1}{\sqrt[4]{g}}
    \frac{\varGamma\left(\frac{5}{4}\right)}
    {\pi}\int_{\mathbb{R}}d\xi A_1(\xi),
\end{equation}
which coincides with the expression (\ref{Phi-4_min}) for the minorant of the $\mathcal{S}$-matrix in the $\varphi^4$ theory.

\section{Conclusion}
\label{sect:conclusion}

Quantum theory with the Lagrangian interaction $\sinh^{2(p+1)}\varphi$ for $p\in\mathbb{N}$, in Euclidean space (the metric signature is all pluses) with arbitrary dimension $D$ was considered in the paper. The expression for the $\mathcal{S}$-matrix of a theory at the zero value of the classical field (argument of the $\mathcal{S}$-matrix) of such a theory in terms of the GCPF (PT series) demonstrates a rigid divergence of order $e^{n^2}$, where $n\in\mathbb{N}$ is the order of the PT series term. For this reason, it is important to construct another series in terms of interaction action $S$ powers. To achieve this, three steps were taken, corresponding to the three sections~\ref{sect:gaussian-integrals-intro},~\ref{sect:pert-theory} and ~\ref{sect:scattering-matrix} of the paper.

In the section~\ref{sect:gaussian-integrals-intro} of the paper a detailed introduction to the theory of Gaussian measures, first for finite-dimensional HS, and then for infinite-dimensional generalization was presented. Such measures were constructed in detail as product measures. This is important because such an approach is rarely presented rigorously in the physical literature. The definition of an integral over such measures in HS was also formulated and the simplest integrals were calculated using MCT and DCT. At the end of the section~\ref{sect:gaussian-integrals-intro}, the formulation of the problem was outlined, in particular, the $\mathcal{S}$-matrix of the theory at the zero value of the classical field was defined in terms of the integral over the Gaussian measure with nuclear covariance operator in the separable GP and linear changes of variables were discussed, leading to equivalent or singular Gaussian measures relative to the original one.

The section~\ref{sect:pert-theory} of the paper is devoted to constructing an expression for the $\mathcal{S}$-matrix at the zero value of the classical field in terms of the GCPF, that is PT series. Despite the strong growth of PT series terms, useful conclusions for further derivation were drawn. In particular, for the theory with the Lagrangian interaction $\sinh^{2(p+1)}\varphi$, using the Mayer cluster expansion, each term of the series was written in terms of Bell polynomials. The number of cluster integrals on which such a polynomial depends can be estimated using Bell numbers.

In the main section~\ref{sect:scattering-matrix} of the paper, an expression was obtained for the $\mathcal{S}$-matrix at the zero value of the classical field in terms of an iterated series in powers of $S$. The first multiplication by ``smart one'' of the integrand yielded the expansion of the exponential function in an orthogonal system of polynomials in reciprocal powers of the new action of the theory. After which the validity of summation and integration permutation was proven, that is not possible in ordinary PT series. By the second multiplication by ``smart one'' of the integrand the expansion of each integrand into the system of generalized Laguerre polynomials depending on positive powers of the new action of the theory, was obtained. After this, the validity of summation and integration permutation was again proven, probably the most difficult problem in the presented paper. At the second step of the $\mathcal{S}$-matrix calculation, it was possible to express the latter in terms of an iterated series (it is important that such a series doesn't reduce to a multiple series) of integrals of the powers of $S$ multiplied by an additional Gaussian exponential function with positive argument over the Gaussian measure. The critical value of exponential function argument was found and ``critical mass'' was determined, below which the integrals converge and are again expressed in terms of the Bell polynomials already studied above.

The convergence of the new series is also determined by the fact that the new propagator that appears during the derivation contains a ``polarization operator'', which suppresses the exponential function argument for large values of the summation index. This leads to behavior $e^{n}$ that is no longer dominant in the coefficients of the resulting series. Let us note that technically useful difference between the obtained result and the result given by traditional PT is that the interaction constant can be transferred to the denominator, leaving the interaction Lagrangian in the numerator, which provides advantages for numerical calculations. At the end of the section~\ref{sect:scattering-matrix}, as a complementary result, asymptotics were obtained for the $\mathcal{S}$-matrix for large values of the coupling constant $g$. These asymptotics, along with the resulting series, completely describe the behavior of the $\mathcal{S}$-matrix at the zero value of the classical field. The development of the methods used in the paper and the results obtained in the paper, in particular, the use of the hard-sphere gas and stochastic cluster approximations, is the subject of future research and publication.

In conclusion, let us note that instead of expanding the integrals over Gaussian measure into convergent series, it is possible to study the question of their approximation by a finite number of an absolutely convergent series terms. In other words, is it possible, with any given accuracy, to approximate the integral over the Gaussian measure by a finite sum depending on this accuracy? The answer to this question for scalar field theories with polynomial interactions turns out to be positive, and the proof is given in two papers~\cite{BelokurovI,BelokurovII} that deserve special mention. These papers inspired our own research.
\section*{Acknowledgments}

The authors are deeply grateful to their families for their love, wisdom and understanding. We also express special gratitude to Mikhail G. Ivanov at the Department of Theoretical Physics MIPT for reading the paper and many valuable comments as well as Oleg A. Zagriadskii and Stanislav S. Nikolaenko at the Department of Higher Mathematics MIPT for helpful discussions on various mathematical issues raised in the paper. Finally, we are grateful to MIPT students Igor O. Amvrosov, Elizaveta D. Kovalenko and Lev D. Melentev for discussing certain issues of measure theory and their enthusiasm in the topic.

\end{document}